\documentclass{aa}

\usepackage{graphicx}

\usepackage{natbib}
\usepackage{txfonts}
\usepackage{booktabs}
\usepackage[hidelinks]{hyperref}
\usepackage[switch]{lineno}
\usepackage{subcaption}

\newcommand{\Fermi}{{Fermi}}

\begin{document}

   \title{Identifying potentially missed extended sources in the {\Fermi}-LAT 4FGL Catalog using clustering analysis}
    \titlerunning{Clustering analysis of Fermi-LAT unassociated sources in the 4FGL}

   \author{G. Cozzolongo\inst{1},
          A. M. W. Mitchell\inst{1},
          S. T. Spencer\inst{1,2}\thanks{Now at Cherenkov Telescope Array Observatory, Science Data Management Centre (SDMC), Platanenallee 6, 15738 Zeuthen, Germany},
          D. V. Malyshev\inst{1},
          T. Unbehaun\inst{1,3}
          }
    \authorrunning{Cozzolongo et al.}

   \institute{Friedrich-Alexander-Universität Erlangen-Nürnberg, Erlangen Centre for Astroparticle Physics, Nikolaus-Fiebiger-Str. 2, 91058 Erlangen, Germany.
              \email{giovanni.cozzolongo@fau.de}
              \and
              Department of Physics, Clarendon Laboratory, Parks Road, Oxford, OX1 3PU, United Kingdom
              \and
              Max-Planck-Institut f\"ur Kernphysik, Saupfercheckweg 1, 69117 Heidelberg, Germany
             }

   \date{Published 9 September 2026}

    \abstract
    {Since its launch in 2008, the {\Fermi} Large Area Telescope (LAT) has detected thousands of sources, many of which remain unassociated. Some may be extended sources represented in the catalog by multiple point-like entries. The reinterpretation of HESS~J1813$-$178 as a single extended source motivates a systematic search for further missed extended sources.}
    {We search for clusters of unassociated {\Fermi}-LAT sources and test whether single extended-source models describe them better than multiple catalog sources.}
    {We applied the algorithm called density-based spatial clustering of applications with noise (DBSCAN) to 4FGL sources with a linking scale of $\varepsilon = 0.3^\circ$ over $5\,\mathrm{GeV}$--$1\,\mathrm{TeV}$. Each cluster contains at least one unassociated source and up to one extended or point-like associated source from selected categories, including pulsars, pulsar wind nebulae, and supernova remnants. Using \texttt{Fermipy}, we compared extended- and multiple-source models and characterized each candidate spectrally and morphologically, focusing on the Galactic plane.}
    {We identified 48 clusters containing 124 sources, each with at least one unassociated source. For all 8 clusters passing our quality selection, an extended-source model is statistically preferred. At linking scales of $0.4^\circ$ and $0.5^\circ$, all 8 clusters retain their core sources. Cross-matches with the Second Fermi Galactic Extended Sources Catalog (2FGES) and the HESS Galactic Plane Survey (HGPS) showed that 5 clusters overlap known extended sources. The other 3 clusters, including one cluster with a morphology dependent on the interstellar emission model, have no counterpart in either catalog and are new extended-source candidates.}
    {Spatial clustering combined with likelihood-based model comparison can uncover extended sources missed in the {\Fermi}-LAT catalog and complements existing searches.}


   \maketitle

\nolinenumbers

\section{Introduction}
\label{sec:introduction}
Since its launch in 2008, the {\Fermi} Gamma-ray Space Telescope \citep{2009ApJ...697.1071A} has detected thousands of gamma-ray sources through its primary instrument, the Large Area Telescope (LAT). The LAT covers the energy range from $20\,\mathrm{MeV}$ to more than $300\,\mathrm{GeV}$ with a field of view of approximately 2.4\,sr, surveying the entire sky every three hours.

The Fourth {\Fermi}-LAT source catalog \citep{Abdollahi2020,2023arXiv230712546B}, in its Data Release 4 version (4FGL-DR4), covers 14 years of data (August 2008 to August 2022) and lists 7194 sources. Only 82 of these are classified as spatially extended, although many Galactic source classes, such as supernova remnants (SNRs), pulsar wind nebulae (PWNe), molecular clouds illuminated by cosmic rays, and stellar clusters, are intrinsically extended at GeV energies \citep{2017ApJ...843..139A, 2024arXiv241107162A}.

Several searches for extended sources have been carried out with different data selections and strategies. \citet{2012ApJ...756....5L} tested every point-like 2FGL catalog source for spatial extension using two years of Pass\,7\footnote{Pass denotes the version of the data processing algorithms used for {\Fermi}-LAT event reconstruction and analysis.} data, regardless of association status, and identified 25 extended sources. Using six years of Pass\,8 data above 10\,GeV, \citet{2017ApJ...843..139A} focused on the inner Galactic plane ($|b| < 7^\circ$), seeding the search from 3FGL catalog sources and newly detected point sources, and reported 46 extended sources. At high latitudes ($|b| > 7^\circ$), \citet{2018ApJS..237...32A} searched for extended sources using 7.5 years of Pass\,8 data above 1\,GeV, excluding unassociated low-significance sources from their initial background model so that potentially missed point sources would not mask underlying extended emission. This approach yielded 24 extended sources.

Despite these advances, 2065 sources in the 4FGL-DR4 catalog lack firm associations with known astrophysical objects. Some clusters of catalog sources might even be single extended sources, as demonstrated in the case of HESS~J1813$-$178 \citep{2018ApJ...859...69A}. A similar issue has been reported in the Centaurus~A field, where several unassociated sources near the radio lobes might be residuals of incomplete modeling of the extended lobe emission \citep{2023MNRAS.523.4455D}. To address this systematically, we applied density-based spatial clustering of applications with noise (DBSCAN) to the spatial distribution of {\Fermi}-LAT sources, focusing on unassociated sources and those likely to include extended objects. This allowed us to determine whether these clusters are better described by their individual catalog entries or by single extended sources.

Section~\ref{sec:data_and_methods} describes the {\Fermi}-LAT data, the regions of interest (ROIs), sources, background models, and our method for identifying and analyzing potential extended sources. Section~\ref{sec:results} presents our results. Section~\ref{sec:discussion} places our findings in the context of the known Galactic gamma-ray source population. Section~\ref{sec:conclusion} summarizes our conclusions.

\section{Data and methods}
\label{sec:data_and_methods}
 This study has two primary tasks. We first clustered the gamma-ray sources from the 4FGL-DR4 catalog and then performed a morphological and spectral analysis of the identified clusters.

\subsection{Clustering analysis}
\label{subsec:clustering_analysis}

\begin{figure}
\centering
\includegraphics[width=\columnwidth]{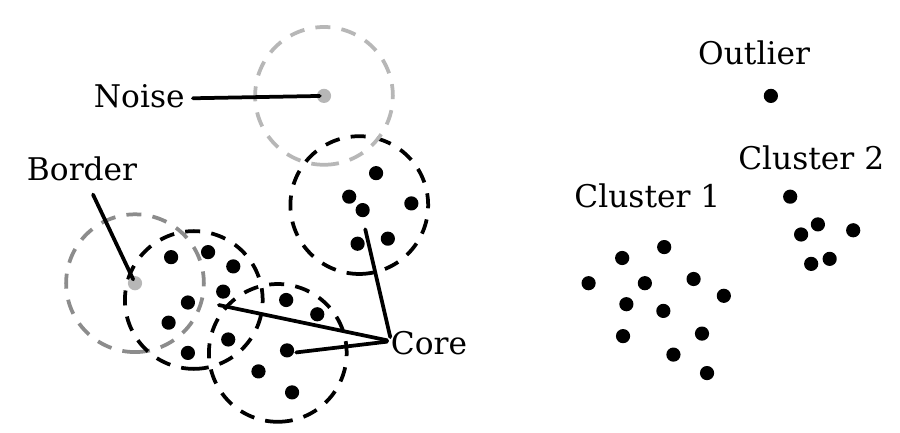}
\caption{ Three point types defined by DBSCAN, with the linking length $\varepsilon$ shown as the circle radius, and MinPts $= 5$. Our analysis used MinPts $= 2$ (Sect.~\ref{subsec:clustering_analysis}), so the minimum cluster size is two points.}
\label{fig:dbscan_scheme}
\end{figure}

The algorithm for our source-clustering analysis was DBSCAN \citep{1996kddm.conf..226E}. DBSCAN is well suited to this task because it does not require the user to specify the number of clusters a priori and naturally separates clustered from isolated sources (noise). We also considered HDBSCAN \citep{10.1007/978-3-642-37456-2_14}, a hierarchical extension of DBSCAN that adaptively determines cluster boundaries without a fixed linking scale. However, for our application, the fixed $\varepsilon$ parameter of DBSCAN is advantageous because it maps directly onto the angular scale of the extended sources we aim to identify (see Fig.~\ref{fig:dbscan_scheme} and the calibration below). The hierarchical approach of HDBSCAN, while more flexible, removes this direct physical interpretability of the linking parameter. DBSCAN operates based on two main parameters: $\varepsilon$, the maximum distance between two points for one to be considered in the neighborhood of the other, and MinPts, the minimum number of points (including the point itself) required in a neighborhood for a point to qualify as a core point. The algorithm creates a circle of radius $\varepsilon$ around every point and classifies each point into one of three categories. Core points have at least MinPts$-1$ neighbors within $\varepsilon$. Border points lie within $\varepsilon$ of a core point but have fewer than MinPts$-1$ neighbors. Noise points are neither core nor border. With MinPts $= 2$, a cluster consists of at least two points: the core point and at least one other point within a distance $\varepsilon$ (see Fig.~\ref{fig:dbscan_scheme}).

\begin{figure}
\centering
\includegraphics[width=\columnwidth]{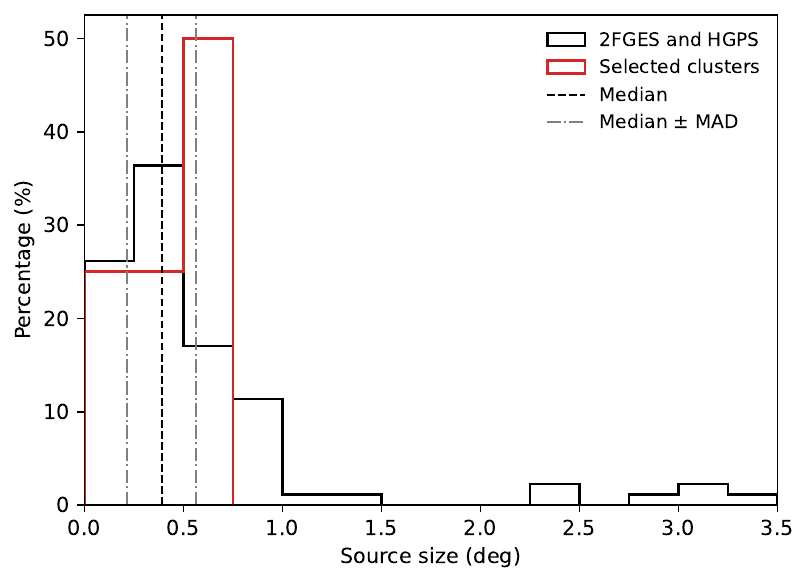}
\caption{Size distribution of extended sources from the combined 2FGES and HGPS catalogs (black) and fitted $r_{68}$ of the eight selected clusters (red). The histograms are independently normalized to $100\%$. The dashed line shows the catalog median ($0.39^\circ$). The dotted lines show the median $\pm$ MAD range. The chosen clustering scale $\varepsilon = 0.3^\circ$ lies within this range.}
\label{fig:size_distribution}
\end{figure}

In our analysis, we set $\varepsilon = 0.3^\circ$ and MinPts $= 2$. We chose $\varepsilon$ based on the size distributions of extended sources reported by \citet{2024arXiv241107162A} for the Second {\Fermi} Galactic Extended Sources Catalog (2FGES) and by \citet{2018A&A...612A...1H} for the HESS Galactic Plane Survey (HGPS), as shown in Fig.~\ref{fig:size_distribution}. Since most of our clusters lie in the Galactic plane, 2FGES provides the most relevant LAT sample of extended GeV sources for setting the clustering scale. The main 4FGL catalog is driven by point sources and imports its extended sources from earlier studies, while some 2FGES sources were released after 4FGL-DR4 and are not yet incorporated there. The joint 2FGES--HGPS calibration is not intended to assume a single emission channel: for leptonic sources, the TeV size can be smaller than the GeV size because higher-energy electrons cool more rapidly, whereas for hadronic sources, the morphology is expected to be more similar across energies \citep{2004vhec.book.....A}. The combined distribution has a median $r_{68} = 0.39^\circ$ and a median absolute deviation (MAD) $0.17^\circ$. Our $\varepsilon = 0.3^\circ$ sits within the median $\pm$ MAD range. We stress that $\varepsilon$ is a linking length and not a source size: DBSCAN chains neighbors transitively, meaning that when source A is linked to source B and source B is linked to source C, all three are assigned to the same cluster even when A and C are separated by more than $\varepsilon$. The algorithm can therefore build clusters much larger than $\varepsilon$.

The MinPts value of two defines the minimum cluster size. We also required that each retained cluster contained at least one unassociated source, as our goal was to investigate whether such sources might be part of extended structures. This requirement was applied a posteriori to the DBSCAN output: DBSCAN was run on the selected source sample, and clusters containing no unassociated source were then discarded.
For the clustering, we included unassociated sources together with sources classified in the 4FGL-DR4 catalog as young pulsars (PSRs), millisecond pulsars (MSPs), pulsar wind nebulae (PWNe), supernova remnants (SNRs), SNR/PWN associations (SPPs), globular clusters (GLCs), star-forming regions (SFRs), or unknown (UNK).
The associated sources in these classes were retained because they might be spatially coincident with extended structures. The requirement that each cluster contains at least one unassociated source ensured that the search targeted potentially misclassified regions. Sources cataloged as spatially extended in 4FGL-DR4 (denoted by the suffix e) were included in the clustering using their catalog positions. Although these sources are already modeled as extended in the catalog, they might represent only partial descriptions of larger emission regions. Our analysis tested whether multiple catalog entries, including both point-like and already-extended sources, are better described collectively by a single extended-source model.
All sources with high-confidence extragalactic associations were excluded from the clustering analysis. However, we retained sources with low-confidence extragalactic classifications, as these might represent misclassified Galactic sources: BL~Lac objects (BLLs), flat-spectrum radio quasars (FSRQs), radio galaxies (RDGs), other active galactic nuclei (AGNs), and blazar candidates of uncertain type (BCU). Retaining them ensured that any cluster member with a tentative extragalactic label was not inadvertently excluded before the likelihood analysis could test whether the emission is actually extended. The remaining minority extragalactic classes (SSRQ, CSS, NLSY1, and SEY) and potentially extended extragalactic classes (GAL and SBG) were excluded, as our analysis targeted Galactic extended sources.

Our analysis yielded 48 distinct clusters, with a total of 124 4FGL-DR4 sources. Each cluster includes at least one unassociated source. There are 36 clusters of size two, 7 clusters of size three, 2 clusters of size five, and 3 clusters of size seven. A complete listing of all clustered sources is provided in Appendix~\ref{appendix:members}. We investigate systematic effects of the $\varepsilon$ choice in Appendix~\ref{appendix:clustering_radius}.

\begin{figure*}
\centering
\includegraphics[width=\textwidth]{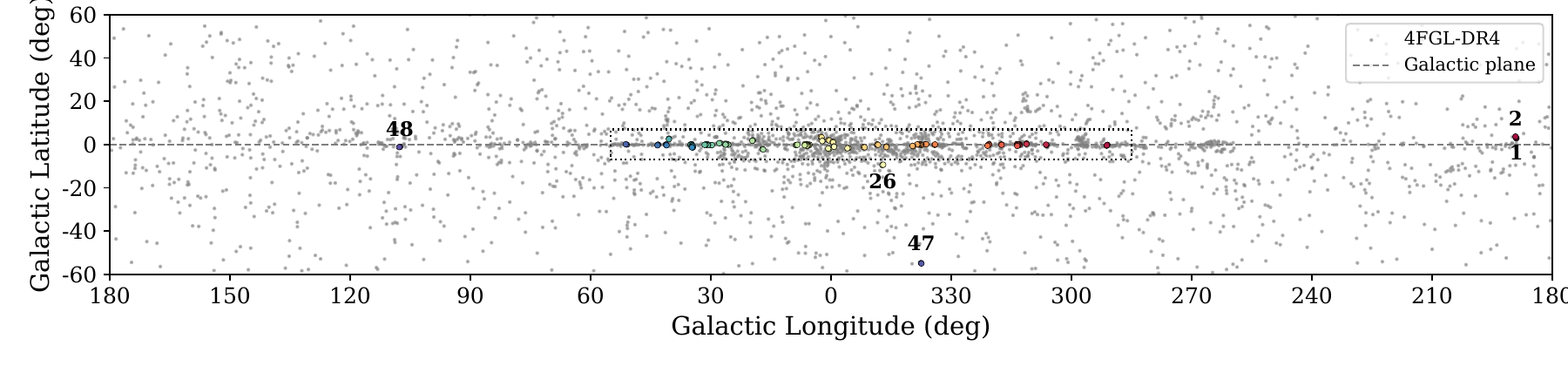}
\caption{{\Fermi}-LAT cluster map. The spatial distribution of the 48 clusters identified in our analysis is overlaid on the Galactic plane. The gray points show all 4FGL-DR4 sources used for the clustering. Each cluster is represented by a different color. A detailed view of the clusters within the dotted rectangle is provided in Fig.~\ref{fig:clusters_hess}, where they are also overlaid on the HGPS contours.}
\label{fig:clusters_lat}
\end{figure*}

\begin{figure*}
\centering
\includegraphics[width=\textwidth]{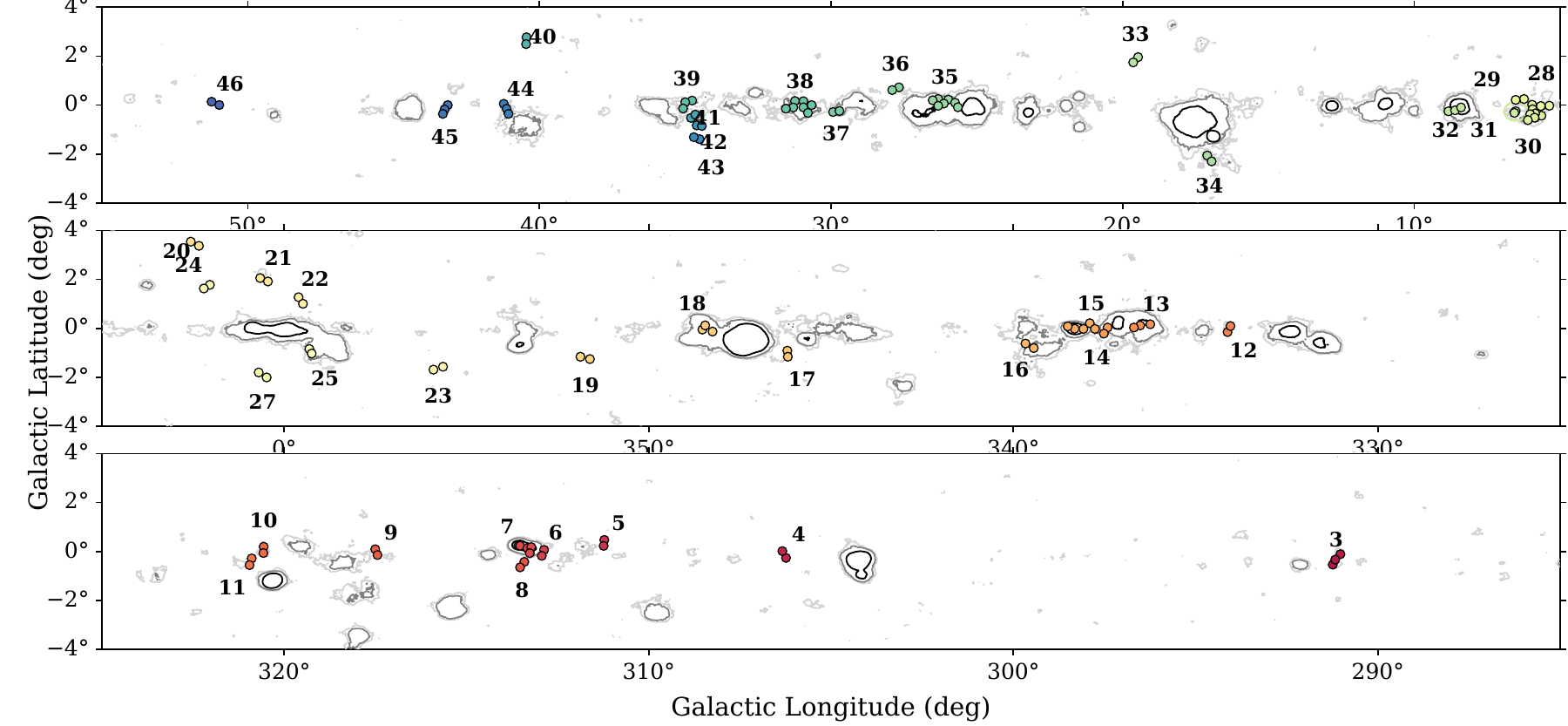}
\caption{Contour map for HGPS. Our identified clusters are overlaid with HGPS contours at 3, 5, and $15\sigma$. Each colored circle represents a single 4FGL-DR4 source belonging to a cluster. The cluster colors match those in Fig.~\ref{fig:clusters_lat}. Only clusters within the dotted rectangle in Fig.~\ref{fig:clusters_lat} are shown.}
\label{fig:clusters_hess}
\end{figure*}

Figure~\ref{fig:clusters_lat} shows the spatial distribution of the clusters identified in our analysis. Figure~\ref{fig:clusters_hess} shows the same clusters overlaid with contours from the HGPS. The overlap shows potential associations between our GeV clusters and known TeV sources.

\subsection{\Fermi-LAT analysis}
\label{subsec:Test_for_extension}
\begin{figure*}
\centering
\includegraphics[width=\textwidth]{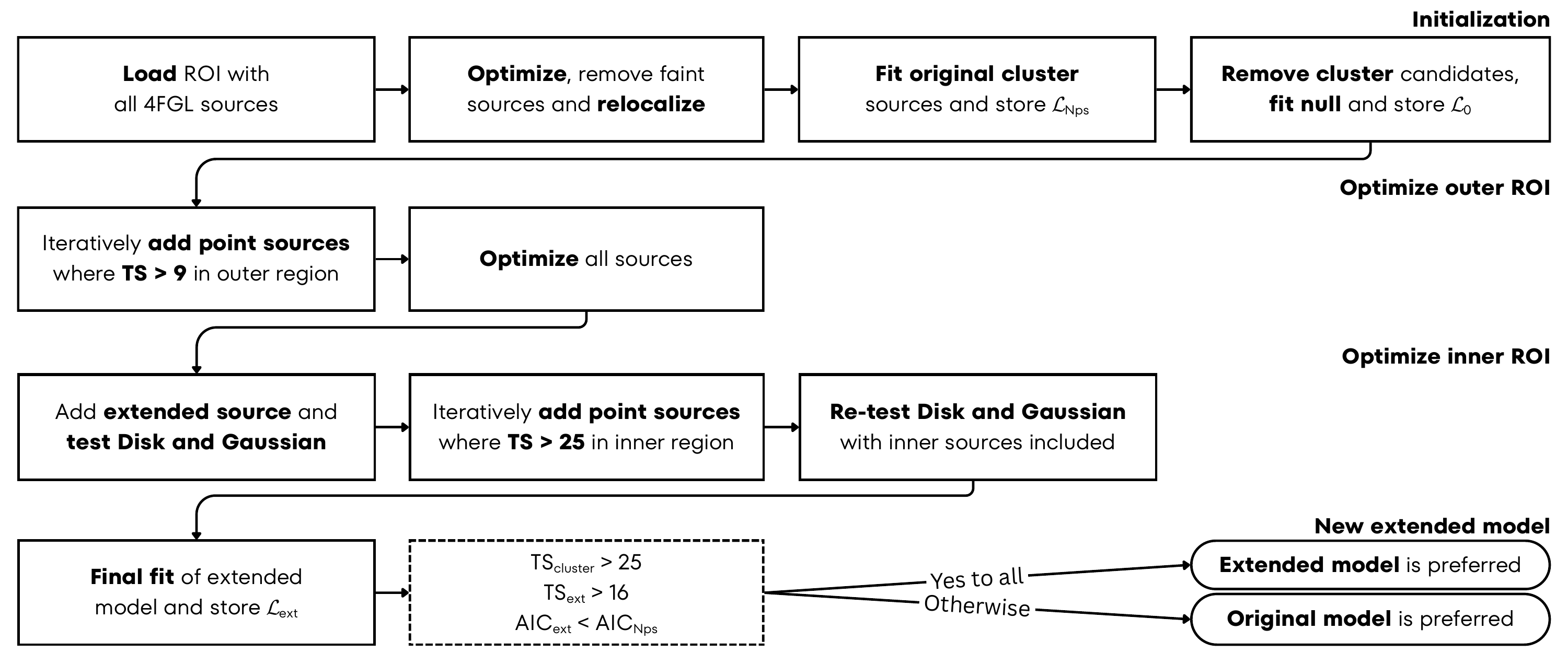}
\caption{Flow chart of the analysis procedure. The solid boxes represent the processing steps, and the dashed boxes indicate the model selection criteria. The inner region is defined as twice the cluster angular size (i.e., twice the maximum angular separation between cluster members), while the outer region extends from this boundary to the ROI edge. The entire procedure was performed independently for each of the three interstellar emission models (Sect.~\ref{subsec:Test_for_extension}).}
\label{fig:diagram}
\end{figure*}

After the clusters were identified by DBSCAN, we performed a likelihood analysis of each cluster using \texttt{Fermipy} (v1.2), which provides a Python interface to the {\Fermi} Science Tools \citep{2017ICRC...35..824W}. Each cluster was analyzed independently. For each cluster, we compared two hypotheses: emission modeled as the individual catalog sources and emission modeled as a single extended source.
An overview of the analysis procedure is shown in Fig.~\ref{fig:diagram}. We performed a joint likelihood analysis over the four point-spread function (PSF) event types (PSF0--PSF3), which partition events by the quality of their reconstructed direction, from broadest (PSF0) to narrowest (PSF3). Each event type was fitted simultaneously with its own instrument response function. We selected the P8R3\_SOURCEVETO event class, which provides reduced cosmic-ray contamination at the cost of a slightly smaller effective area compared to the standard SOURCE class, making it well suited to extended-source searches where background rejection is critical~\citep{2018arXiv181011394B}. We set the minimum energy to $5\,\mathrm{GeV}$ to achieve a PSF lower than $0.1^\circ$ for the highest-quality events (PSF3). At 5\,GeV, the PSF for the lowest-quality events (PSF0) is considerably broader. However, the joint PSF likelihood analysis accounted for this difference by assigning each event type its own instrument response function. We set the maximum energy to $1\,\mathrm{TeV}$ to achieve an energy dispersion below $15\%$. We used data collected from 2008 October 27 to 2022 August 2 (see Table~\ref{tab:data_selection}). We applied the recommended standard cuts and filters, including a zenith-angle cut of $105^\circ$ to avoid gamma-ray contamination from Earth's limb. We adopted the binned maximum-likelihood method to derive the spectral and morphological parameters of the sources. To model the background, we included the sources in the ROI listed in the 4FGL-DR4 catalog along with the recommended Galactic and isotropic diffuse emission components, described by the files \texttt{gll\_iem\_v07.fits} and \texttt{iso\_P8R3\_SOURCEVETO\_V3\_PSF*\_v1.txt}, respectively, and provided with \texttt{Fermitools} \citep{2017ICRC...35..824W}. The energy dispersion correction was applied as recommended by the {\Fermi}-LAT Collaboration.

We used spatial bins of $0.025^\circ$ and performed a binned maximum-likelihood analysis with ten logarithmic bins per decade in energy. Table~\ref{tab:data_selection} summarizes the parameters of our data selection\footnote{For details, see the {\Fermi}-LAT Collaboration's caveats about analyzing LAT Pass 8 (P8R3\_V3) data at \url{https://fermi.gsfc.nasa.gov/ssc/data/analysis/LAT_caveats.html}.}.

\begin{table}[ht]
\caption{Summary of {\Fermi}-LAT data selection criteria.}
\label{tab:data_selection}
\centering
\small
\begin{tabular}{ll}
\hline\hline
Selection & Criteria \\
\hline
Observation period & 2008 October 27 (18:11:14 UTC)\\
 & to 2022 August 2 (08:23:00 UTC) \\
Mission elapsed time (s)\tablefootmark{a} & 246823875 to 681121385 \\
Energy range & $5\,\mathrm{GeV}$--$1\,\mathrm{TeV}$ \\
ROI width & $7\degr$ \\
Source ROI width & $13\degr$ \\
Zenith angle cut & $105\degr$ \\
Data quality cut & \texttt{DATA\_QUAL} $> 0$ \\
 & \texttt{LAT\_CONFIG} $== 1$ \\
Event class & \texttt{P8R3\_SOURCEVETO} \\
Event types & FRONT + BACK \\
 & PSF0, PSF1, PSF2, PSF3 \\
Models & \\
\hspace{1em}Catalog & \texttt{gll\_psc\_v34.fit} \\
\hspace{1em}Galactic diffuse & \texttt{gll\_iem\_v07.fits} \\
IRFs & \texttt{P8R3\_SOURCEVETO\_V3\_PSF0\_v1} \\
 & \texttt{P8R3\_SOURCEVETO\_V3\_PSF1\_v1} \\
 & \texttt{P8R3\_SOURCEVETO\_V3\_PSF2\_v1} \\
 & \texttt{P8R3\_SOURCEVETO\_V3\_PSF3\_v1} \\
\hline
\end{tabular}
\tablefoot{
\tablefoottext{a}{{\Fermi} mission elapsed time is defined as seconds since 2001 January 1, 00:00:00 UTC.}
}
\end{table}

We constructed the initial sky model for each cluster using a $7^\circ$ ROI, containing all 4FGL sources except those with a test statistic (TS) $< 4$ and fewer than five predicted counts in our energy band. A source ROI width of $13^\circ$ was used to account for PSF spillover from sources outside the $7^\circ$ region, following the standard method established in previous {\Fermi}-LAT extended-source studies \citep{2017ApJ...843..139A, 2018ApJS..237...32A}.
After optimizing spectral parameters using the \texttt{optimize} method, we removed faint sources satisfying $\mathrm{TS} < 4$ and fewer than five predicted counts, reoptimized, and relocalized point sources with $\mathrm{TS} > 16$ using the \texttt{localize} method. Our analysis then proceeded in distinct phases, inspired by \citet{2018ApJ...859...69A}.

\paragraph{Initialization.}
We fit all cluster sources simultaneously using \texttt{fit()}, with the spectral freedom of each source determined by its TS value from the preceding optimization step. For $\mathrm{TS} < 4$, only the normalization was let free. For $4 \leq \mathrm{TS} < 100$, the normalization and spectral index were let free. For $\mathrm{TS} \geq 100$, all spectral parameters except the energy scale were let free. We also let free the normalizations of nearby catalog sources within $1.5^\circ$ of the mean position (arithmetic mean of the member coordinates) of the cluster members with $\mathrm{TS} \geq 9$\footnote{This choice limits the number of free parameters to prevent overfitting and numerical instabilities.} and the diffuse background components. The fit was iterated up to 20 times and terminated early when $\Delta \ln \mathcal{L} < 0.1$ between consecutive iterations. This provided the original-model likelihood $\mathcal{L}_\mathrm{Nps}$. During this initialization fit, the parameters were counted as effectively fitted only when they were let free and included in a successful fit, defined as a fit where the \texttt{Fermipy} fit-quality flag was $= 3$ (on a scale of 0 to 3, where 3 indicates full convergence). When a fit returned quality $< 3$, it was retried after the spectral parameters were frozen (beyond the normalization) to their catalog values. Only the parameters that remained free in the successful retry were counted as effectively fitted. Parameters that were freed but were subsequently reset to their catalog values because all retries also fail were excluded from the count, as they do not contribute to the descriptive power of the model. We then removed all clustered sources from the model whose 4FGL-DR4 classification was unassociated or belonged to a source class that might be spatially extended, regardless of whether the association confidence was high (uppercase CLASS1) or low (lowercase CLASS1): supernova remnants (snr/SNR), pulsar wind nebulae (pwn/PWN), supernova remnant / pulsar wind nebula associations (spp/SPP), star-forming regions (sfr/SFR), and globular clusters (glc/GLC). Clustered sources with a high-confidence CLASS1 entry in a class that was not expected to be spatially extended (e.g., identified pulsars) were kept in the model, and only their normalization was allowed to vary. We fitted this null model to obtain $\mathcal{L}_0$.

\paragraph{Optimize the outer ROI.} We searched for residual sources in the outer ROI (farther than twice the cluster angular size from the center) by iteratively generating TS maps with a test source with a spectral index of 2 and adding a point source at the position of the most significant peak, provided its $\mathrm{TS}$ exceeded $9$, requiring a minimum separation of $0.5^\circ$ from any existing model source (4FGL catalog sources and previously added residual sources), until no more significant peaks remained.

\paragraph{Optimize the inner ROI.}
We added an extended source at the mean position of the cluster members and tested the two spatial models available in \texttt{Fermipy} (symmetric disk and symmetric Gaussian), with the radius, position, and spectral parameters left free. We selected the model with the highest $\mathrm{TS}_\mathrm{ext}$ as the preliminary best model. We generated a TS map and iteratively added point sources where peaks within the inner ROI (within twice the cluster angular size from the center) exceeded $\mathrm{TS} > 25$, requiring a minimum separation of $0.5^\circ$ from any existing model source (4FGL catalog sources, the center of the extended source, and any previously added residual sources). The higher threshold for the inner ROI was used because we already modeled the main emission with an extended source, so any remaining residuals must be highly significant to warrant additional point sources. When inner sources were added, we retested the two spatial models (disk and Gaussian) with the inner sources included in the model and selected the model with the highest $\mathrm{TS}_\mathrm{ext}$. Any point sources that were added in this step remained in the model for all subsequent likelihood evaluations ($\mathcal{L}_\mathrm{ext}$, $\mathcal{L}_\mathrm{1ps}$), ensuring that the model comparison accounted for all significant emission components within the cluster region.

\paragraph{New extended model.} We performed a final fit of the extended model with up to 20 iterations, applying the same convergence criterion ($\Delta \ln \mathcal{L} < 0.1$) as in the initialization. This provided the extended-model likelihood $\mathcal{L}_\mathrm{ext}$.

The TS maps shown in Appendix~\ref{appendix:tsmaps} were generated after the extended-source fitting procedure. To produce each map, we started from the null model (i.e., without the cluster sources and without the extended source). The removed cluster sources were those whose classification was unassociated or belonged to a potentially extended source class (shown as cyan crosses in the figures).  Point sources with high-confidence classifications (e.g., identified pulsars with a CLASS1 entry) that happen to be spatially coincident with the cluster were retained in the model because they represent known astrophysical objects. The TS map was then computed by placing a test point source with a spectral index of 2 at each pixel position and calculating $\mathrm{TS} = 2\ln(\mathcal{L}_{\mathrm{test}}/\mathcal{L}_{\mathrm{null}})$, where $\mathcal{L}_{\mathrm{test}}$ is the likelihood with the test source and $\mathcal{L}_{\mathrm{null}}$ is the likelihood without it.

The likelihood values were defined as follows. $\mathcal{L}_0$ is the likelihood for the null model with all cluster sources removed. $\mathcal{L}_\mathrm{ext}$ is the likelihood for the model with a single extended source at the cluster position. $\mathcal{L}_\mathrm{1ps}$ is the likelihood for the extended-source model with the spatial extension set to zero, reducing it to a single point source. Finally, $\mathcal{L}_\mathrm{Nps}$ is the likelihood for the original 4FGL model retaining the $N$ catalog cluster sources, which can include sources cataloged as extended in 4FGL-DR4.

We defined several TS values to quantify the model comparisons: the detection test statistic, $\mathrm{TS}_\mathrm{cluster} = 2 \ln \left( \mathcal{L}_\mathrm{ext} / \mathcal{L}_{0} \right)$, measures the significance of the extended source relative to a null hypothesis with no source at the cluster position (i.e., all cluster sources are removed from the model). The source-extension test statistic is given by $\mathrm{TS}_\mathrm{ext} = 2 \ln \left( \mathcal{L}_\mathrm{ext} / \mathcal{L}_\mathrm{1ps} \right)$, and it quantifies the preference for an extended model over a single point-source model. Finally, the $N$-source test statistic is $\mathrm{TS}_\mathrm{Nps} = 2 \ln \left( \mathcal{L}_\mathrm{Nps} / \mathcal{L}_{0} \right)$, which quantifies the significance of the $N$ catalog sources relative to the null model. In accordance with \citet{2017ApJ...843..139A}, we applied the following criteria: a detection was claimed for sources with $\mathrm{TS} > 25$, corresponding to a significance of approximately $4\sigma$ for a single source \citep{1996ApJ...461..396M}. To define a source as extended, we used a threshold of $\mathrm{TS}_\mathrm{ext} > 16$, corresponding to a significance of nearly $4\sigma$ for extension.

However, the significance of non-nested models (e.g., comparing an extended-source model to a multiple-source model) cannot be quantitatively compared using a simple likelihood ratio test. We therefore also considered the Akaike information criterion (AIC), introduced by \citet{Akaike1974} and defined as
\begin{equation}
\mathrm{AIC} = 2k - 2\ln(\mathcal{L}),
\end{equation}
where $k$ is the number of free parameters, and $\mathcal{L}$ is the likelihood.
The best model is the one that minimizes the AIC. Comparing the AIC for extended and $N$-source models leads to
\begin{equation}
\mathrm{AIC}_\mathrm{ext} < \mathrm{AIC}_\mathrm{Nps} \iff \mathrm{TS}_\mathrm{ext} > \mathrm{TS}_\mathrm{Nps} + 2\Delta k,
\end{equation}
where $\Delta k = k_\mathrm{ext} - k_\mathrm{Nps}$ is the difference in the number of free parameters (extended model minus $N$-source model).

The total parameter count $k$ includes all free parameters in the likelihood fit: (i) spectral parameters of the sources under investigation (cluster sources or extended source plus any residual inner point sources), (ii) normalization parameters of nearby catalog sources with $\mathrm{TS} \geq 9$, (iii) spectral parameters of any sources added in the outer ROI during the iterative source-finding procedure, and (iv) parameters of the diffuse background components (Galactic and isotropic).

For the individual cluster sources, we counted spectral parameters (normalization, spectral index, and curvature or cutoff energy for curved spectra), plus two positional parameters (RA, Dec) for each source that was relocalized successfully (sources not meeting the $\mathrm{TS} > 16$ relocalization threshold retained their catalog positions and contribute no positional degrees of freedom). For extended sources, we counted two positional parameters and one extension parameter (radius or $\sigma$, depending on the spatial model) from the extension scan, plus the spectral parameters from the final fit.

Common parameters, specifically, nearby catalog sources, outer sources added during the optimization of the outer ROI, and diffuse background components, are identical between the $N$-source and extended-source models by construction because they share the same ROI optimization up to and including the outer-ROI step. These cancel exactly in $\Delta\mathrm{AIC}$, so only the cluster-source parameters contribute to $\Delta k$. Since the ROI optimization and outer-source search were performed independently for each interstellar emission model (IEM), the number of outer sources $N_\mathrm{outer}$ can differ across models for a given cluster. This was accounted for in the $\Delta\mathrm{AIC}$ comparison, which was computed separately within each IEM realization.

As described in the inner-ROI optimization step, we selected the spatial model (disk or Gaussian) with the highest $\mathrm{TS}_\mathrm{ext}$ \citep{2018ApJS..237...32A}.
In addition, we performed detailed spectral analyses of the candidate extended sources. We considered all three spectral models available in \texttt{Fermipy}: a simple power law, log parabola, and a power law with an exponential cutoff.
We used the \texttt{sed()} method to compute the spectral energy distribution (SED) for the individual cluster sources and the extended-source model, adopting a fixed power-law index of $\Gamma = 2$ within each energy bin (\texttt{bin\_index\,=\,2.0}). The power-law model is defined as
\begin{equation}
\frac{dN}{dE} = N_0 \left(\frac{E}{E_0}\right)^{-\Gamma},
\end{equation}
where $N_0$ is the prefactor, $\Gamma$ is the spectral index, and $E_0$ is the energy scale. While we tested multiple spectral models (power law, log parabola, and exponential cutoff) for all sources during the ROI optimization, all statistically significant extended sources in our final sample are well described by simple power-law spectra. No cluster shows a statistically significant preference ($\mathrm{TS}_\mathrm{curve} > 16$) for curved spectral models over simple power laws.

\begin{figure*}[htbp]
\centering
\begin{minipage}[b]{0.45\textwidth}
    \centering
    \includegraphics[width=\textwidth]{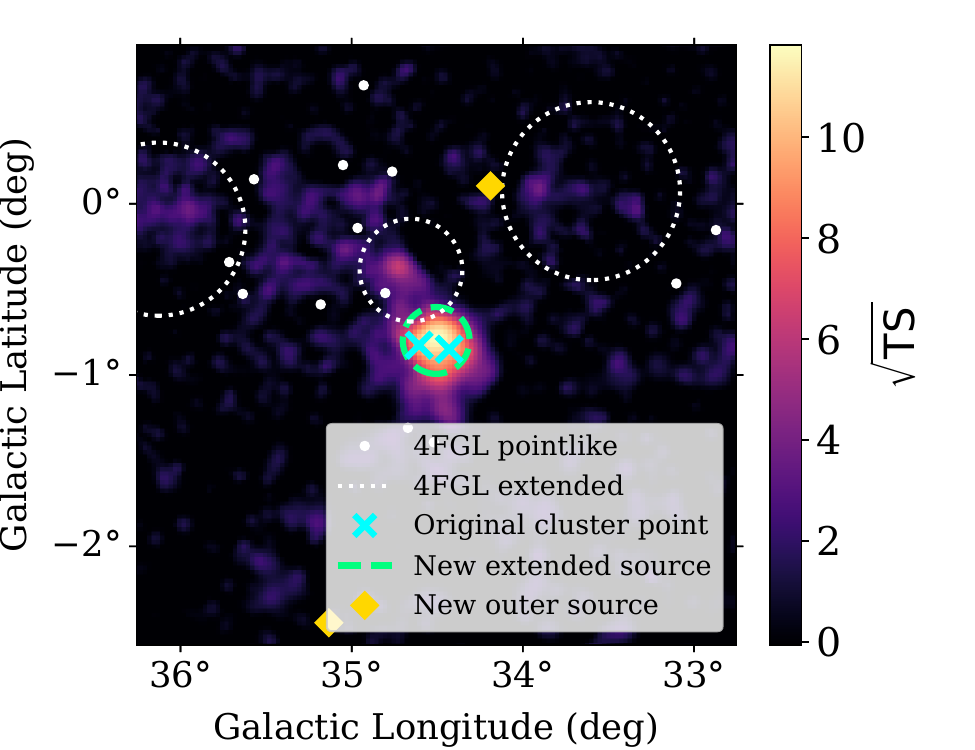}
\end{minipage}
\hspace{0.5cm}
\begin{minipage}[b]{0.457\textwidth}
    \centering
    \includegraphics[width=\textwidth]{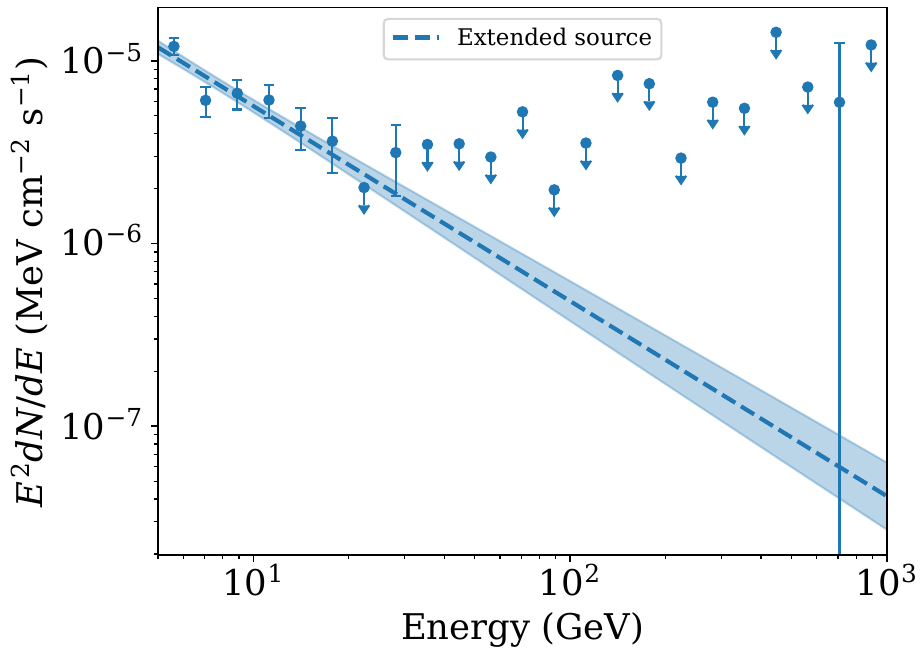}
\end{minipage}
\vspace{0.5cm}
\begin{minipage}[b]{0.45\textwidth}
    \centering
    \includegraphics[width=\textwidth]{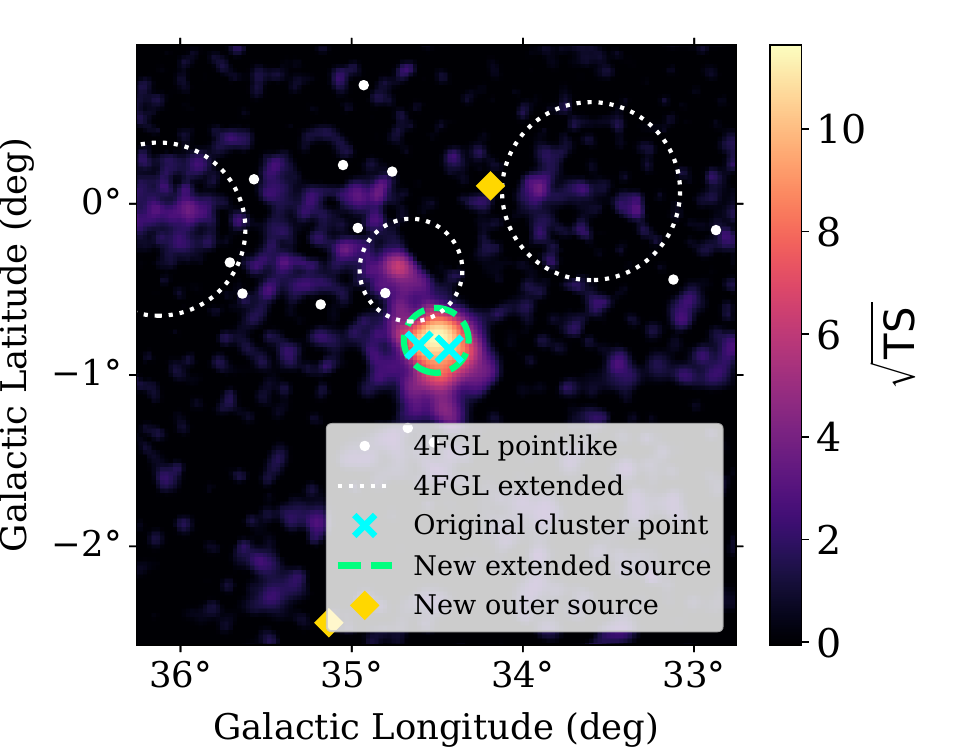}
\end{minipage}
\hspace{0.5cm}
\begin{minipage}[b]{0.457\textwidth}
    \centering
    \includegraphics[width=\textwidth]{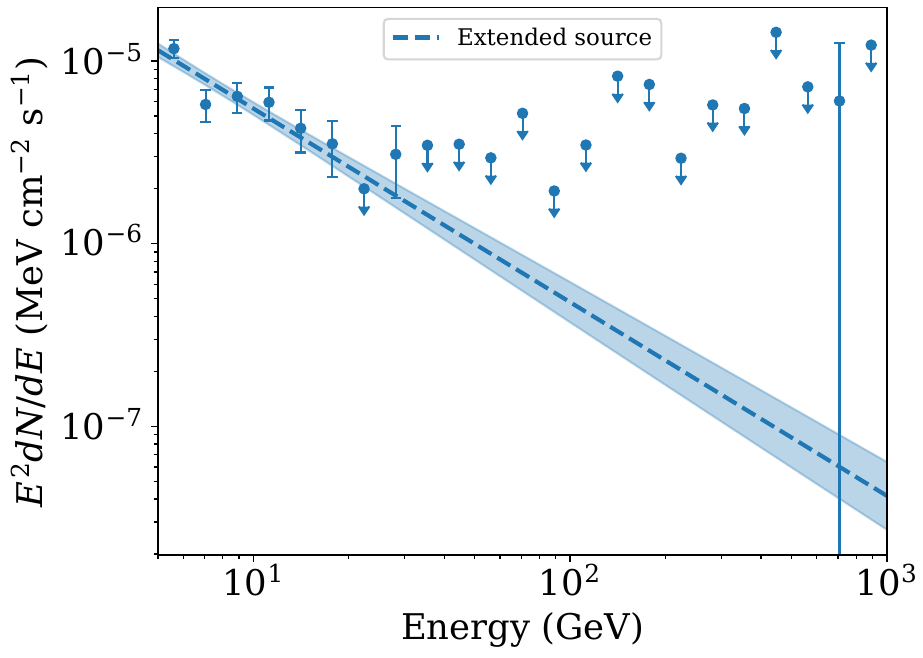}
\end{minipage}
\vspace{0.5cm}
\begin{minipage}[b]{0.45\textwidth}
    \centering
    \includegraphics[width=\textwidth]{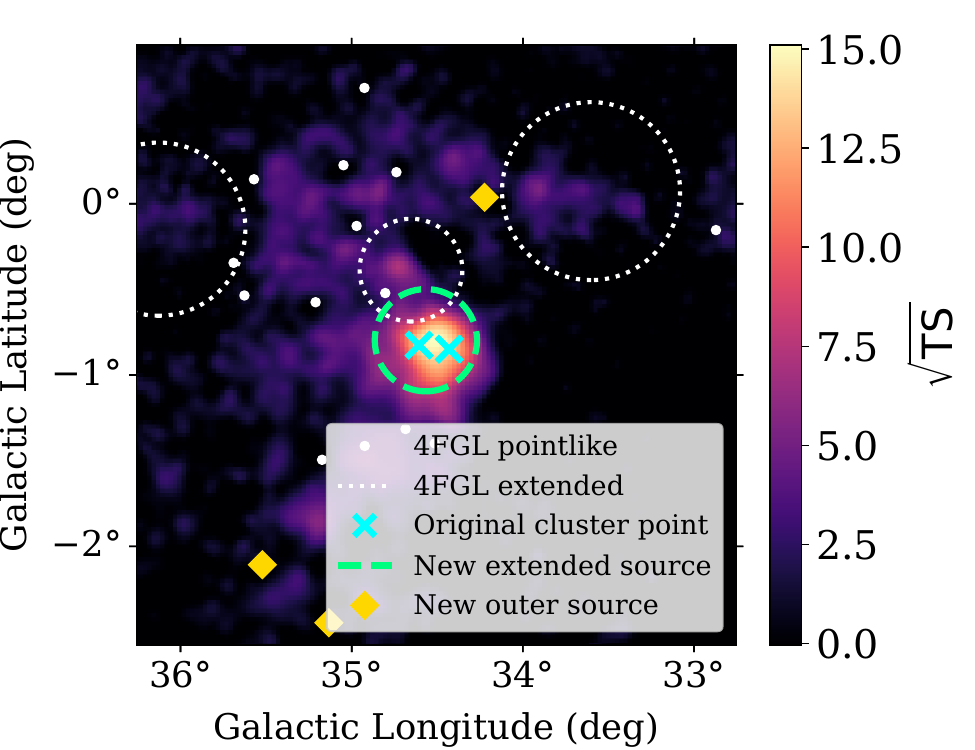}
\end{minipage}
\hspace{0.5cm}
\begin{minipage}[b]{0.457\textwidth}
    \centering
    \includegraphics[width=\textwidth]{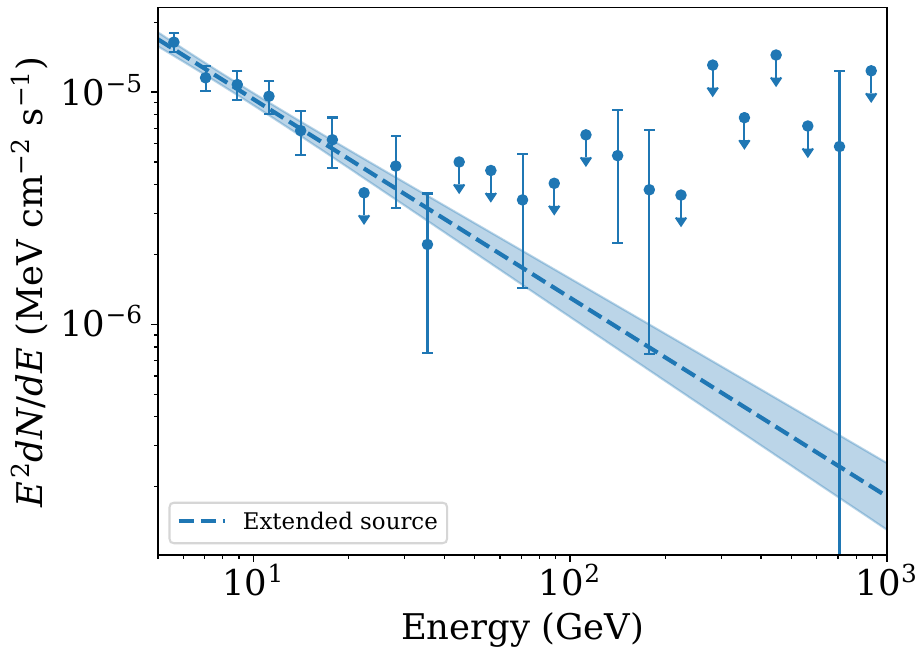}
\end{minipage}
\caption{TS maps (left) and SEDs (right) of cluster~42 for the standard, unpatched, and \texttt{GALPROP} background models, shown from top to bottom. In the TS maps, white dots represent 4FGL point-like sources. White dotted circles show 4FGL extended sources not belonging to the cluster, while cyan dotted circles show 4FGL extended sources that are cluster members. Cyan crosses mark original cluster sources that are unassociated or classified with lower confidence and were therefore removed from the model. The green dashed line shows the $68\%$ containment radius for the extended source under investigation. Positions of 4FGL point-like sources may differ slightly between rows because source relocalization is performed independently for each background model during the ROI optimization process (see Sect.~\ref{subsec:Test_for_extension}). In the SEDs, upper limits are shown for flux points with significance lower than $2\sigma$. }
\label{fig:cluster_42_combined}
\end{figure*}

For each cluster, this analysis was performed using three different diffuse emission models to test the robustness of our detections. The first was the {\Fermi}-LAT Galactic diffuse emission model, IEM v07 \citep{2016ApJS..223...26A}, which incorporates various components to account for gamma-ray emission from interstellar processes. These components include an ad hoc patch component, which is a set of spatial templates added to absorb large-scale residual emission that is not well represented by the physics-based interstellar emission model, including the Galactic center excess, loop~I, and the {\Fermi} bubbles. As our second diffuse emission model, we used the unpatched model, which is identical to the standard model, but without the patch described above. The unpatched IEM was included as a uniform stress test on the diffuse model across the sample. The patch templates act on angular scales above about $2^\circ$, while our clusters have typical sizes of $0.3^\circ$--$0.5^\circ$, so the patches are not expected to drive the diffuse model for most clusters. This comparison also served as a cautionary check for applications of the method at larger linking lengths, for which the cluster scale approaches the patch scale (Appendix~\ref{appendix:clustering_radius}). The third model was generated using \texttt{GALPROP} \citep{2022ApJS..262...30P}, which is a numerical code for calculating the propagation of relativistic charged particles and the diffuse emissions produced during their propagation. An example with the three emission models is shown in Fig.~\ref{fig:cluster_42_combined}.

Although the {\Fermi}-LAT IEM \citep{2016ApJS..223...26A} is also based on \texttt{GALPROP} for the Galactic emission components, there are many adjustments in the {\Fermi}-LAT IEM compared to the original \texttt{GALPROP} prediction, including a rescaling of atomic and molecular gas $\pi^0$ components in energy bins and in galactocentric annuli, addition of components describing dark neutral medium (DNM) and dust-corrected atomic hydrogen, and adjustment of the inverse-Compton component in each energy bin. In our \texttt{GALPROP} model, we only modulated the total $\pi^0$ and IC components by power-law functions with free normalization and spectral index correction (tilt). There are also some differences in the \texttt{GALPROP} parameters. In particular, we used 2D diffusion with a maximum radius $R_\mathrm{max} = 20\,\mathrm{kpc}$ and height $h_\mathrm{max} = 10\,\mathrm{kpc}$ and a parameterized distribution of CR sources \citep{2011CoPhC.182.1156V}, while the model of \citet{2016ApJS..223...26A} used CR sources proportional to the distribution of pulsars in the Galaxy \citep{2004A&A...422..545Y}, $R_\mathrm{max} = 30\,\mathrm{kpc}$, and $h_\mathrm{max} = 6\,\mathrm{kpc}$. Our systematic uncertainty estimate from these alternative models is likely conservative because the models represent relatively extreme deviations from the standard IEM.

For certain source types, the spatial extent can exhibit an energy dependence. In particular, for sources dominated by energy-loss processes such as pulsar wind nebulae, the emission region can become more compact at higher energies because the highest-energy particles cool faster. However, for sources with geometrically defined boundaries such as supernova remnants, this energy dependence can be less pronounced or absent. Testing for energy-dependent morphology is beyond the scope of this work and is left for future dedicated analyses of individual sources.

\section{Results}
\label{sec:results}
Our analysis of the {\Fermi}-LAT data using the DBSCAN algorithm revealed a total of 48 distinct clusters, listed in Table~\ref{tab:clusters_all}. These clusters encompass 124 sources cataloged in the 4FGL-DR4, each containing at least one previously unassociated source. Only 8 of them, listed in Table~\ref{tab:clusters_significant}, satisfy all three selection criteria ($\mathrm{TS}_\mathrm{cluster} > 25$, $\mathrm{TS}_\mathrm{ext} > 16$, and $\Delta\mathrm{AIC} > 0$) and have no critical quality flags (b, c, or d), as defined in Table~\ref{tab:clusters_all}. Flag~(a), indicating IEM-dependent spatial model preference, was not treated as a rejection criterion but is discussed separately. The remaining 40 clusters were rejected by the selection criteria or flagged for unreliable fits. The fit results are shown in Table~\ref{tab:clusters_fit_results} and the TS maps for the standard emission model in Appendix~\ref{appendix:tsmaps}. The cluster associations with FGES and HGPS catalog sources are listed in Tables~\ref{tab:fges_associations} and \ref{tab:hgps_associations}, respectively. The circle overlap percentages were calculated using the $68\%$ containment radii of our fitted sources and the catalog sources. The conversion from model parameters into $r_{68}$ is given in the footnote to Table~\ref{tab:clusters_fit_results}. We used the published $r_{68}$ values directly for 2FGES and HGPS catalog sources.

Following \citet{2016ApJS..224....8A}, we quantified the overlap between source regions using two complementary parameters. The positional overlap, $O_\mathrm{loc} = A_\cap / \min(A_1, A_2)$, equals unity when the smaller region is fully contained in the larger one. The extension overlap, $O_\mathrm{ext} = A_\cap / \max(A_1, A_2)$, penalizes size mismatches and equals unity only when the two regions are identical. A source pair is matched when $O_\mathrm{loc} > 0.4$ and $O_\mathrm{ext} > 0.4$, and it is marginal when both exceed $0.1$, but at least one is below $0.4$.

\begin{table}
\caption{\label{tab:clusters_significant}Statistical test results for selected extended-source candidates.}
\centering
\scriptsize
\setlength{\tabcolsep}{2.2pt}
\begin{tabular}{lrrrrrr}
\hline
ID & $\mathrm{TS}_\mathrm{Nps}$ & $\mathrm{TS}_\mathrm{cluster}$ & $\mathrm{TS}_\mathrm{ext}$ & $k_\mathrm{Nps}$ & $k_\mathrm{ext}$ & $\Delta\mathrm{AIC}$ \\
\hline\hline
3 & 60 & 129 & 31 & 12 & 18 & 57 \\
7 & 416 & 693 & 211 & 17 & 34 & 243 \\
13 & 1280 & 1442 & 1028 & 9 & 18 & 144 \\
24 & 54 & 179 & 24 & 7 & 19 & 101 \\
30 & 784 & 921 & 462 & 17 & 23 & 125 \\
32 & 1236 & 1278 & 815 & 7 & 16 & 24 \\
36 & 97 & 159 & 38 & 9 & 18 & 44 \\
42 & 237 & 358 & 80 & 19 & 24 & 111 \\
\hline
\end{tabular}
\tablefoot{$\mathrm{TS}_\mathrm{Nps}$: combined TS of the original $N$ catalog sources.
$\mathrm{TS}_\mathrm{cluster}$: TS of extended-source model.
$\mathrm{TS}_\mathrm{ext}$: extension significance (extended vs. single point source).
$k_\mathrm{Nps}$, $k_\mathrm{ext}$: number of free parameters for the $N$-source and extended models.
$\Delta\mathrm{AIC} = \mathrm{AIC}_\mathrm{Nps} - \mathrm{AIC}_\mathrm{ext}$
(positive favors extended model).
$\mathrm{TS}_\mathrm{ext}$ compares extended vs.\ single point source.
$\Delta\mathrm{AIC}$ compares $N$ catalog sources vs.\ extended source.}
\end{table}

\begin{table*}
\caption{\label{tab:clusters_fit_results}Spatial and spectral fit results for selected extended-source candidates.}
\centering
\small
\renewcommand{\arraystretch}{1.3}
\begin{tabular}{lcccccc}
\hline
ID & Model & $l$ & $b$ & $r_{68}$ & Prefactor & $\Gamma$ \\
 & & (deg) & (deg) & (deg) & ($\mathrm{cm^{-2}\,s^{-1}\,MeV^{-1}}$) & \\
\hline\hline
3 & G & $291.17 \pm 0.03^{+0.01}$ & $-0.36 \pm 0.05^{+0.09}$ & $0.53 \pm 0.07_{-0.46}$ & $(3.4 \pm 0.4_{-2.4}) \times 10^{-14}$ & $2.39 \pm 0.11^{+0.12}$ \\
7 & D & $313.38 \pm 0.01_{-0.03}$ & $0.22 \pm 0.02_{-0.13}$ & $0.34 \pm 0.01^{+0.04}$ & $(6.0 \pm 0.5^{+2.0}) \times 10^{-14}$ & $1.93 \pm 0.05^{+0.08}$ \\
13 & G & $336.48 \pm 0.01^{+0.08}$ & $0.10 \pm 0.01_{-0.07}$ & $0.71 \pm 0.02^{+0.02}$ & $(2.2 \pm 0.1^{+0.1}_{-0.1}) \times 10^{-13}$ & $2.05 \pm 0.03_{-0.04}$ \\
24 & D & $2.14 \pm 0.03$ & $1.67 \pm 0.02_{-0.01}$ & $0.11 \pm 0.02_{-0.02}$ & $(9.2 \pm 1.6^{+0.7}_{-5.1}) \times 10^{-15}$ & $2.13 \pm 0.13^{+0.05}_{-0.17}$ \\
30 & G & $5.98 \pm 0.01^{+0.06}$ & $-0.39 \pm 0.01^{+0.02}$ & $0.52 \pm 0.02^{+0.26}$ & $(1.5 \pm 0.1^{+1.1}) \times 10^{-13}$ & $2.33 \pm 0.05^{+0.03}_{-0.01}$ \\
32 & G & $8.47 \pm 0.01$ & $-0.11 \pm 0.01_{-0.03}$ & $0.50 \pm 0.02^{+0.05}$ & $(1.7 \pm 0.1^{+0.5}) \times 10^{-13}$ & $2.07 \pm 0.04^{+0.06}_{-0.01}$ \\
36 & D & $27.77 \pm 0.02^{+0.10}$ & $0.68 \pm 0.02_{-0.05}$ & $0.14 \pm 0.02_{-0.14}$ & $(2.0 \pm 0.3^{+0.1}_{-1.2}) \times 10^{-14}$ & $2.74 \pm 0.15^{+0.88}_{-0.06}$ \\
42 & G & $34.51 \pm 0.02^{+0.06}$ & $-0.80 \pm 0.02$ & $0.29 \pm 0.03^{+0.16}_{-0.01}$ & $(5.7 \pm 0.5^{+3.7}_{-0.2}) \times 10^{-14}$ & $3.07 \pm 0.11_{-0.21}$ \\
\hline
\end{tabular}
\tablefoot{Model: G (Gaussian) or D (disk).
$r_{68}$: radius containing $68\%$ of the flux, computed as
$\sigma\sqrt{-2\ln 0.32}$ for Gaussian and $R\sqrt{0.68}$ for disk models.
Prefactor quoted at a reference energy $E_0 = 10\,\mathrm{GeV}$.
$\Gamma$: power-law spectral index.
All central values from the standard IEM fit.
Uncertainties: value $\pm$ stat $^{+\sigma_\mathrm{syst}}_{-\sigma_\mathrm{syst}}$,
where systematic errors are the difference between the standard IEM value and
max/min across three IEM models.
Systematic errors are defined as $\sigma_\mathrm{syst}^{+} = x_\mathrm{max} - x_\mathrm{std}$ and $\sigma_\mathrm{syst}^{-} = x_\mathrm{std} - x_\mathrm{min}$, where $x_\mathrm{std}$ is the standard IEM value, and $x_\mathrm{max}$, $x_\mathrm{min}$ are the maximum and minimum across the three IEM models. When $x_\mathrm{std} = x_\mathrm{min}$ (i.e.\ $\sigma_\mathrm{syst}^{-} = 0$), only the upper systematic is shown. When $x_\mathrm{std} = x_\mathrm{max}$ (i.e.\ $\sigma_\mathrm{syst}^{+} = 0$), only the lower systematic is shown. When both are nonzero, both are reported.
Analysis energy range: $5\,\mathrm{GeV}$ -- $1\,\mathrm{TeV}$.
The systematic uncertainties reported here were derived solely from the three interstellar emission models and do not account for uncertainties in the instrument response functions (IRFs).}
\end{table*}

\begin{table*}
\caption{\label{tab:fges_associations}Spatial cross-match between selected clusters and 2FGES catalog sources.}
\centering
\small
\begin{tabular}{lccccccl}
\hline\hline
ID & 2FGES source & $r_{68}^\mathrm{cluster}$ & $r_{68}^\mathrm{2FGES}$ & Distance & $O_\mathrm{loc}$ & $O_\mathrm{ext}$ & Match \\
 & & (deg) & (deg) & (deg) & (\%) & (\%) & \\
\hline
3 & J1112.6$-$6059 & $0.53 \pm 0.07_{-0.46}$ & $0.39 \pm 0.06$ & $0.074 \pm 0.067^{+0.059}$ & $100.0 \pm 0.0$ & $53.8 \pm 22.3^{+0.5}_{-50.5}$ & matched \\
7 & J1417.5$-$6057 & $0.34 \pm 0.01^{+0.04}$ & $0.17 \pm 0.04$ & $0.196 \pm 0.032_{-0.006}$ & $96.1 \pm 12.0^{+3.9}$ & $23.6 \pm 9.1_{-4.1}$ & marginal \\
13 & J1633.5$-$4743 & $0.71 \pm 0.02^{+0.02}$ & $0.57 \pm 0.03$ & $0.131 \pm 0.051_{-0.059}$ & $100.0 \pm 0.0$ & $63.6 \pm 7.7_{-3.4}$ & matched \\
30 & J1759.4$-$2355 & $0.52 \pm 0.02^{+0.26}$ & $0.83 \pm 0.06$ & $0.300 \pm 0.054^{+0.007}$ & $100.0 \pm 0.0_{-19.3}$ & $39.6 \pm 6.9^{+31.3}_{-0.7}$ & marginal \\
\hline
\end{tabular}
\tablefoot{All central values are from the standard IEM fit. Uncertainties are given as value $\pm$ stat $^{+\sigma_\mathrm{syst}}_{-\sigma_\mathrm{syst}}$. $r_{68}$: $68\%$ containment radius.
Distance: angular separation between the cluster and 2FGES source centers.
$O_\mathrm{loc}$: positional overlap $= A_\cap / \min(A_1, A_2)$.
$O_\mathrm{ext}$: extension overlap $= A_\cap / \max(A_1, A_2)$,
following \citet{2016ApJS..224....8A}.
A source pair is matched when $O_\mathrm{loc} > 0.4$ and $O_\mathrm{ext} > 0.4$.
A source pair is marginal when both $> 0.1$, but at least one $< 0.4$. It is unmatched otherwise.
2FGES $r_{68}$ values and errors from \citet{2024arXiv241107162A}.}
\end{table*}

\begin{table*}
\caption{\label{tab:hgps_associations}Spatial cross-match between selected clusters and HGPS catalog sources.}
\centering
\small
\begin{tabular}{lcccccccl}
\hline\hline
ID & HGPS source & Class & $r_{68}^\mathrm{cluster}$ & $r_{68}^\mathrm{HGPS}$ & Distance & $O_\mathrm{loc}$ & $O_\mathrm{ext}$ & Match \\
 & & & (deg) & (deg) & (deg) & (\%) & (\%) & \\
\hline
7 & J1418$-$609 & PWN & $0.34 \pm 0.01^{+0.04}$ & $0.163 \pm 0.016$ & $0.164 \pm 0.017_{-0.033}$ & $100.0 \pm 0.0$ & $22.4 \pm 4.6_{-4.6}$ & marginal \\
7 & J1420$-$607 & PWN & $0.34 \pm 0.01^{+0.04}$ & $0.122 \pm 0.010$ & $0.199 \pm 0.014^{+0.086}$ & $100.0 \pm 0.0_{-5.1}$ & $12.6 \pm 2.1_{-3.1}$ & marginal \\
13 & J1632$-$478 & Unid & $0.71 \pm 0.02^{+0.02}$ & $0.275 \pm 0.030$ & $0.180 \pm 0.026^{+0.100}$ & $100.0 \pm 0.0$ & $14.8 \pm 3.3_{-0.8}$ & marginal \\
13 & J1634$-$472 & Unid & $0.71 \pm 0.02^{+0.02}$ & $0.263 \pm 0.020$ & $0.661 \pm 0.020_{-0.053}$ & $59.1 \pm 7.3^{+16.1}$ & $8.0 \pm 1.3^{+1.8}$ & unmatched \\
30 & J1800$-$240 & Unid & $0.52 \pm 0.02^{+0.26}$ & $0.478 \pm 0.058$ & $0.038 \pm 0.041^{+0.059}$ & $100.0 \pm 0.0$ & $83.5 \pm 21.9^{+1.5}_{-45.8}$ & matched \\
30 & J1801$-$233 & SNR & $0.52 \pm 0.02^{+0.26}$ & $0.257 \pm 0.045$ & $0.684 \pm 0.014_{-0.061}$ & $10.7 \pm 6.5^{+73.1}_{-0.9}$ & $2.6 \pm 2.2^{+6.5}_{-0.2}$ & unmatched \\
32 & J1804$-$216 & Unid & $0.50 \pm 0.02^{+0.05}$ & $0.379 \pm 0.034$ & $0.089 \pm 0.056^{+0.017}$ & $100.0 \pm 0.0$ & $56.8 \pm 10.8^{+0.2}_{-9.2}$ & matched \\
\hline
\end{tabular}
\tablefoot{All central values are from the standard IEM fit. Uncertainties are given as value $\pm$ stat $^{+\sigma_\mathrm{syst}}_{-\sigma_\mathrm{syst}}$. Point-like HGPS sources are excluded. Some HGPS sources have no uncertainty on $r_{68}$. In these cases only the central value is reported. $r_{68}$: $68\%$ containment radius.
For HGPS Gaussian sources, $r_{68} = \sigma\sqrt{-2\ln 0.32}$.
For shell and multi-Gaussian morphologies, the catalog $R_{70}$ is used as approximation.
Distance: angular separation between cluster and HGPS source centers.
$O_\mathrm{loc}$: positional overlap $= A_\cap / \min(A_1, A_2)$.
$O_\mathrm{ext}$: extension overlap $= A_\cap / \max(A_1, A_2)$,
following \citet{2016ApJS..224....8A}.
A source pair is matched when $O_\mathrm{loc} > 0.4$ and $O_\mathrm{ext} > 0.4$.
A source pair is marginal when both $> 0.1$, but at least one $< 0.4$. It is unmatched otherwise.
Class: HGPS classification (PWN, SNR, composite, binary, unid).}
\end{table*}

The morphological and spectral fit results for the eight selected clusters are summarized in Table~\ref{tab:clusters_fit_results}. All eight clusters are located within $|b| < 3^\circ$, consistent with Galactic stellar remnants and star-forming regions.

To test the sensitivity of our clustering to the choice of $\varepsilon$, we reran DBSCAN at $0.4^\circ$ and $0.5^\circ$ (Appendix~\ref{appendix:clustering_radius}). Only the cluster membership was recomputed. The full likelihood analysis was not repeated. All eight selected clusters retained their core sources at all tested scales, although many expanded or merged at larger radii (Table~\ref{tab:radius_analysis_summary}).

\section{Discussion}
\label{sec:discussion}

The high rejection rate of the DBSCAN candidates (40/48, or 83\%) shows that most spatial groupings are chance alignments and that the selection criteria are stringent enough to keep the false-positive rate under control. The DBSCAN step is intentionally permissive (it flags any spatial grouping of unassociated sources regardless of physical plausibility), and the likelihood analysis acts as the filter. Clusters of cataloged point sources can trace real extended emission for several reasons. Source confusion is the most straightforward: the finite angular resolution of the LAT can split distinct emission peaks within a single extended structure into separate catalog entries, as in the case of HESS~J1813$-$178 \citep{2018ApJ...859...69A}. Another example is the Centaurus~A region, where unassociated catalog sources around the radio lobes have been interpreted as likely residuals of imperfect extended-source modeling \citep{2023MNRAS.523.4455D}. The catalog pipeline might also decompose diffuse emission from supernova remnants, pulsar wind nebulae, or pulsar halos into multiple point-like detections. In some regions, physically distinct but spatially overlapping sources can contribute emission that is collectively better captured by a single extended model at GeV energies.

The source confusion described above accounts for the large significance jumps when clusters of catalog sources are replaced by extended models. If several weak marginally detected 4FGL sources are fragments of a single brighter extended source, the fitting of a single spatial template recovers flux that was distributed among multiple low-significance components, producing a much stronger detection.

If the preference for extended models is physical and not an artifact of diffuse background modeling, some 4FGL point sources might trace substructure within larger emission regions and do not represent independent objects. We tested this by repeating the analysis with three different diffuse emission models. Clusters that survived across all three are more likely to be genuine, whereas model-dependent detections should be treated with caution.

As demonstrated in Appendix~\ref{appendix:clustering_radius}, the number and complexity of identified clusters grow rapidly with increasing $\varepsilon$ (Table~\ref{tab:radius_analysis_summary}), and analyzing all potential extended sources at larger scales requires substantial additional computational resources that are beyond the scope of this initial systematic study.

\begin{table}
\caption{\label{tab:good_cluster_associations}Spatial matching of selected clusters with known extended-source catalogs.}
\centering
\small
\begin{tabular}{lcc}
\hline\hline
Category & $N$ & Cluster IDs \\
\hline
HGPS and 2FGES & 3 & 7, 13, 30 \\
HGPS only & 1 & 32 \\
2FGES only & 1 & 3 \\
No association & 3 & 24, 36, 42 \\
\hline
\end{tabular}
\tablefoot{Clusters are grouped according to whether they overlap with HGPS sources, 2FGES sources, both, or neither. Clusters with no catalog association are new potential extended-source candidates. Associations were determined by the overlap parameters $O_\mathrm{loc}$ and $O_\mathrm{ext}$
of the $r_{68}$ containment regions, following \citet{2016ApJS..224....8A}
(see Tables~\ref{tab:hgps_associations} and~\ref{tab:fges_associations}).
A cluster is associated when at least one match is matched or marginal.}
\end{table}

Table~\ref{tab:good_cluster_associations} groups the eight selected clusters by their spatial association with known extended-source catalogs. Of these, six clusters (clusters 7, 13, 24, 30, 32, and 42) carry no quality flags, while two clusters (clusters 3 and 36) carry flag~(a), indicating that their preferred spatial model differs across IEM realizations. We organize the discussion below accordingly: first the confirmations of known sources and the new candidates of the six clusters without quality flags, and then the two flag~(a) candidates that require additional caution in their morphological interpretation.

To place the eight selected clusters in their multiwavelength context, we cross-matched each cluster region with the SNRcat catalog of Galactic supernova remnants \citep{2012AdSpR..49.1313F}, the ATNF Pulsar Catalogue \citep{2005AJ....129.1993M}, and the SIMBAD astronomical database \citep{2000A&AS..143....9W}. The associations discussed below are drawn from these catalogs.

\paragraph{Confirmations of known extended sources.}
Four clusters overlap spatially with HGPS sources, providing independent GeV confirmations of known TeV-emitting regions. Of these, clusters~7, 13, and 30 also match 2FGES entries, giving them multi-catalog support.

Cluster~7 is marginally associated with HGPS~J1418$-$609 and J1420$-$607 and also with 2FGES~J1417.5$-$6057. This is a complex region around $l \sim 313^\circ$ (Kookaburra region) containing multiple known PWNe and pulsars, where our single-extended-source model does not distinguish overlapping emission components. Dedicated multi-source analyses are required.

Cluster~13 is marginally associated with HGPS~J1632$-$478 and matches 2FGES~J1633.5$-$4743, while the match with HGPS~J1634$-$472 falls below the marginal threshold. The first FGES catalog \citep{2017ApJ...843..139A} includes FGES~J1633.0$-$4746, whose 4FGL counterpart (4FGL~J1633.0$-$4746e) is a cluster member. This source was modeled as a disk in the catalog, but our analysis prefers a Gaussian. The filled-center SNR~G336.4$+$0.2 lies close to the fitted centroid, and the energetic pulsar PSR~J1632$-$4757 provides a plausible powering source for the observed emission. With the largest fitted radius in our sample and a hard spectral index, this cluster traces a well-established extended emission region.

Cluster~30, located in the W~28 region, matches HGPS~J1800$-$240 and is marginally associated with 2FGES~J1759.4$-$2355. The shell-type SNR~G005.7$-$0.1 and W~28 itself (the thermal composite SNR~G006.4$-$0.1) both lie within the cluster region. This is a known complex area where interactions between the SNR shock and nearby molecular clouds produce extended gamma-ray emission. Similar to Cluster~7, this region contains multiple overlapping emission structures that require analyses beyond a single extended-source model.

Cluster~32 matches HGPS~J1804$-$216, but has no counterpart in the 2FGES catalog. However, the first FGES catalog \citep{2017ApJ...843..139A} include FGES~J1804.8$-$2144, which is spatially coincident with this cluster and with a hard spectrum consistent with our measurement. This source exhibits an energy-dependent morphology: above 10\,GeV, the emission is coincident with the TeV source, while at lower energies, it is spatially shifted, as reported by \citet{2017ApJ...843..139A}. A dedicated study by \citet{2019ApJ...881...94L} confirmed this picture, showing that the low-energy {\Fermi}-LAT emission is consistent with the intermediate-age SNR W\,30, while the high-energy component is likely produced by a PWN. The region hosts two energetic pulsars, PSR~J1803$-$2137 and PSR~J1803$-$2149, both with a spin-down luminosity $\sim10^{36}$\,erg, and the shell-type SNR~G008.3$-$0.0 and the thermal and plerionic composite SNR~G008.7$-$0.1, consistent with the HGPS associations for HESS~J1804$-$216.

\paragraph{New extended-source candidates.}
Two clusters without catalog counterparts (clusters 24 and 42) also carry no quality flags, making them the strongest new extended-source candidates in our sample. They are located in the Galactic plane ($|b| < 3^\circ$), consistent with Galactic stellar remnants or star-forming regions. No cataloged pulsars or SNRs fall within the cluster~24 region, but the AGILE source 2AGL~J1743$-$2613c and the X-ray source 1RXS~J174419.9$-$261230 are positionally coincident and plausible counterparts, although also with an unassociated origin. With a hard spectral index ($\Gamma = 2.13 \pm 0.13$) and the most compact morphology ($r_{68} = 0.11 \pm 0.02^\circ$), cluster~24 is a particularly promising target for follow-up observations at TeV energies with current and future Cherenkov telescopes.

The thermal and plerionic composite SNR~G034.7$-$0.4 (W44) is spatially coincident with cluster 42. W44 is already cataloged as an extended source with a ring-like morphology in 4FGL, although this ring structure is not evident in our TS map (Appendix~\ref{appendix:tsmaps}). A recent study by \citet{2025A&A...693A.255A} resolved two distinct extended components in the W44 region: a brighter source to the northwest and a fainter, more compact source to the southeast. Our cluster potentially traces the southeastern component, although our fitted morphology ($r_{68} = 0.29 \pm 0.03^\circ$) is larger than that reported by \citet{2025A&A...693A.255A}. The steep spectral index ($\Gamma = 3.07 \pm 0.11$) is consistent with emission from an evolved interacting SNR. The emission traced by our cluster is likely associated with the ears of W44, which are regions in which cosmic rays that escape from the SNR shell interact with adjacent molecular clouds \citep{2012ApJ...749L..35U}. There are no energetic pulsars in the region.

\paragraph{Candidates carrying flag~(a).}
Of the 8 selected clusters, 2 carry flag~(a), meaning that their preferred spatial model (disk vs.\ Gaussian) differs across the three IEM realizations. These candidates pass all selection criteria, but require additional caution in interpreting their morphology. We also verified the local contribution of the patch templates in the standard IEM. Of the 8 selected clusters, 2 (clusters~24 and 36, both close to the Galactic center) sit in regions in which the patch contribution exceeds $5\%$ of the local diffuse flux, while for the remaining 6 selected clusters, the patch contribution stays below $5\%$. For the full sample, 15 of the 48 clusters have a patch contribution above the same threshold. For all 8 selected clusters, the \texttt{GALPROP} model gives a larger deviation from the standard fit than the unpatched model, and this therefore drives the IEM systematic.

Cluster~3 matches 2FGES~J1112.6$-$6059 and has no HGPS counterpart. The plerionic composite SNR~G291.0$-$0.1 is the most likely counterpart, which has previously been associated with Fermi-LAT GeV sources. While the 2FGES overlap provides evidence of extended GeV emission, the absence of a TeV counterpart leaves its physical nature less constrained. The young massive stellar cluster NGC~3603, for which diffuse gamma-ray emission has been reported by \citet{2017A&A...600A.107Y} and which is also listed as a potential cosmic-ray accelerator by \citet{Mitchell2026}, is positionally coincident and represents an alternative counterpart. The energetic pulsar PSR\,J1112$-$6103, with a spin-down luminosity $4.5\times10^{36}$\,erg/s, lies in the cluster region. Its distance is model dependent, ranging from $\sim4.5$\,kpc based on the NE2001 electron density model \citep{2005AJ....129.1993M} to higher values depending on the adopted model. At the closer distance, PSR\,J1112$-$6103 might be a plausible powering source for the observed emission. A firm assessment requires an independent distance measurement.

Cluster~36 has no spatial counterpart in the HGPS or 2FGES catalogs and is therefore a new candidate, although the IEM-dependent morphology indicates that its fitted shape is not robust against background modeling. Located in the Galactic plane with a moderately steep spectrum, it would benefit from deeper observations and dedicated analyses with additional diffuse emission models before its morphological parameters can be considered reliable. The SNR G027.8+00.6 is a possible counterpart and is already listed as a potential association for the member source 4FGL\,J1840.0-0411. A candidate PWN was detected within G027.8$+$0.6 \citep{2010ApJ...725..931M}. A dedicated analysis is needed to determine whether the gamma-ray emission is associated with the SNR or the candidate PWN.

\section{Conclusion}
\label{sec:conclusion}
We applied DBSCAN clustering to unassociated and potentially extended {\Fermi}-LAT sources in the $5\,\mathrm{GeV}$--$1\,\mathrm{TeV}$ range (SOURCEVETO event class, PSF types 0--3), followed by likelihood-based comparisons of extended-source and multiple-source models. This procedure identified 48 clusters from the 4FGL-DR4 catalog.

We report the fitted parameters for all eight clusters for which the extended-source model is statistically preferred over multiple catalog sources, with no critical quality flags in Table~\ref{tab:clusters_fit_results}. Some of these candidates may warrant inclusion in future LAT catalog releases, although dedicated analyses of individual regions are needed to confirm their nature. More broadly, the clustering approach presented here is complementary to existing extended-source searches and can help reduce source confusion in crowded regions of the Galactic plane.

A natural next step is to investigate individual clusters in their TeV and multiwavelength context. This includes conducting joint analyses of individual ROIs using Fermi-LAT and H.E.S.S.\ data, as already done for the Crab nebula by \citet{2024A&A...686A.308A}, with tools such as \texttt{Gammapy} \citep{2023A&A...678A.157D}.

\begin{acknowledgements}
GC, AM and STS are supported by the Deutsche Forschungsgemeinschaft (DFG, German Research Foundation) -- Project Number 452934793. The work of DM is supported by the DFG grant MA~8279/3-1.

We thank Marianne Lemoine-Goumard, Jean Ballet, and David Smith for their thorough internal review on behalf of the Fermi-LAT Collaboration, which significantly improved this manuscript.

The Fermi LAT Collaboration acknowledges generous ongoing support from a number of agencies and institutes that have supported both the development and the operation of the LAT as well as scientific data analysis. These include the National Aeronautics and Space Administration and the Department of Energy in the United States, the Commissariat \`a l'Energie Atomique and the Centre National de la Recherche Scientifique / Institut National de Physique Nucl\'eaire et de Physique des Particules in France, the Agenzia Spaziale Italiana and the Istituto Nazionale di Fisica Nucleare in Italy, the Ministry of Education, Culture, Sports, Science and Technology (MEXT), High Energy Accelerator Research Organization (KEK) and Japan Aerospace Exploration Agency (JAXA) in Japan, and the K.~A.~Wallenberg Foundation, the Swedish Research Council and the Swedish National Space Board in Sweden.

Additional support for science analysis during the operations phase is gratefully acknowledged from the Istituto Nazionale di Astrofisica in Italy and the Centre National d'Etudes Spatiales in France.

The authors acknowledge the use of the following software:
\texttt{Astropy}~\citep{2013A&A...558A..33A},
\texttt{Matplotlib}~\citep{Hunter:2007},
\texttt{pandas}~\citep{mckinney-proc-scipy-2010},
\texttt{scikit-learn}~\citep{scikit-learn},
\end{acknowledgements}

\section*{Data availability}
The DBSCAN cluster lists obtained for linking lengths of $\varepsilon = 0.3^\circ$, $0.4^\circ$, and $0.5^\circ$ are available in machine-readable form on Zenodo at \url{https://doi.org/10.5281/zenodo.21235066}.

\bibliographystyle{aa}
\bibliography{references}

\onecolumn

\begin{appendix}

\section{Cluster properties and members}
\label{appendix:members}

This appendix lists the cluster-level properties and all 4FGL-DR4 catalog sources comprising each of the 48 clusters identified through our DBSCAN analysis.

\begin{table}[h!]
\caption{\label{tab:clusters_all}Identification, catalog associations, and quality flags for all clusters.}
\centering
\small
\begin{tabular}{lccp{8cm}ccc}
\hline
ID & $N_\mathrm{src}$ & Classes & Associations & Position ($l$, $b$) & Max.\ sep. & Flags \\
 & & & & (deg) & (deg) & \\
\hline\hline
1 & 2 & SNR & IC 443 & (189.01, 2.97) & 0.15 & -- \\
2 & 2 & & & (189.17, 3.57) & 0.29 & d \\
3 & 3 & PSR & PSR J1111-6039, PSR J1112-6103 & (291.14, -0.33) & 0.48 & a \\
4 & 2 & & & (306.29, -0.12) & 0.29 & b,c \\
5 & 2 & & & (311.23, 0.35) & 0.25 & a,b,c \\
6 & 2 & & & (312.90, -0.06) & 0.25 & b,c \\
7 & 5 & PWN, PSR & HESS J1420-607, PSR J1418-6058, PSR J1420-6048 & (313.38, 0.14) & 0.44 & -- \\
8 & 2 & PSR & PSR J1422-6138 & (313.47, -0.54) & 0.25 & a,b,c \\
9 & 2 & spp & & (317.47, -0.03) & 0.24 & b,c \\
10 & 2 & & & (320.57, 0.07) & 0.26 & d \\
11 & 2 & & & (320.92, -0.42) & 0.28 & a,b,c \\
12 & 2 & & & (334.08, -0.03) & 0.26 & a,d \\
13 & 3 & spp, pwn & HESS J1632-478 & (336.49, 0.10) & 0.47 & -- \\
14 & 2 & & & (337.47, -0.09) & 0.27 & a,d \\
15 & 5 & spp & HESS J1640-465, HESS J1641-463 & (338.11, 0.04) & 0.75 & a \\
16 & 2 & PSR & PSR J1648-4611 & (339.55, -0.71) & 0.29 & a,b,c,d \\
17 & 2 & & & (346.19, -1.04) & 0.25 & a,b,c \\
18 & 3 & unk, PSR & 1RXS J171538.1-383531, CTB 37A, PSR J1714-3830 & (348.41, -0.02) & 0.32 & -- \\
19 & 2 & unk & NVSS J172917-364008 & (351.74, -1.21) & 0.28 & a,d \\
20 & 2 & & & (2.45, 3.46) & 0.28 & a,d \\
21 & 2 & & & (0.55, 1.98) & 0.25 & a,d \\
22 & 2 & & & (359.55, 1.14) & 0.29 & a,b,c,d \\
23 & 2 & PSR & PSR J1742-3321 & (355.77, -1.63) & 0.29 & b,c \\
24 & 2 & unk & 1RXS J174419.9-261230 & (2.12, 1.70) & 0.22 & -- \\
25 & 2 & PSR & PSR J1747-2958 & (359.28, -0.94) & 0.20 & a,d \\
26 & 2 & & & (347.10, -9.40) & 0.27 & -- \\
27 & 2 & & & (0.59, -1.91) & 0.30 & a,d \\
28 & 2 & PSR, unk & 1RXS J175835.5-240203, PSR J1757-2421 & (5.52, -0.03) & 0.29 & b,c \\
29 & 2 & unk & NVSS J175948-230944 & (6.38, 0.23) & 0.29 & b,c \\
30 & 7 & unk & 1RXS J175851.9-234511, HESS J1800-240A, HESS J1800-240B, NVSS J180041-235622 & (5.92, -0.35) & 0.64 & -- \\
31 & 2 & SNR & & (6.55, -0.28) & 0.08 & b \\
32 & 3 & unk, SNR, SPP & HESS J1804-216, NVSS J180615-212601 & (8.61, -0.19) & 0.46 & -- \\
33 & 2 & & & (19.56, 1.85) & 0.27 & b,c \\
34 & 2 & & & (17.02, -2.18) & 0.29 & b \\
35 & 7 & PSR, unk & NVSS J183922-055321, PSR J1837-0604, PSR J1838-0537 & (26.10, 0.10) & 0.91 & -- \\
36 & 2 & unk, spp &  & (27.79, 0.67) & 0.26 & a \\
37 & 2 & pwn & & (29.82, -0.26) & 0.22 & a,d \\
38 & 7 & unk, glc & HESS J1848-018, NVSS J184840-020408 & (31.06, -0.05) & 0.89 & a,d \\
39 & 3 & & & (34.94, 0.06) & 0.45 & a,d \\
40 & 2 & unk & NVSS J185612+075340 & (40.45, 2.63) & 0.28 & a,d \\
41 & 2 & SNR & & (34.73, -0.45) & 0.20 & a,d \\
42 & 2 &  &  & (34.52, -0.84) & 0.18 & -- \\
43 & 2 & & & (34.61, -1.35) & 0.21 & d \\
44 & 3 & PSR, snr & PSR J1906+0722 & (41.13, -0.15) & 0.44 & d \\
45 & 3 & snr & & (43.23, -0.18) & 0.39 & -- \\
46 & 2 & SNR & & (51.11, 0.07) & 0.29 & d \\
47 & 2 & MSP & PSR J2241-5236 & (337.50, -54.82) & 0.23 & a,d \\
48 & 2 & spp & & (107.66, -1.12) & 0.28 & b,c \\
\hline
\end{tabular}
\tablefoot{$N_\mathrm{src}$: number of 4FGL-DR4 sources in cluster.
Classes and associations from 4FGL-DR4 (CLASS1, ASSOC1, ASSOC\_TEV columns).
Max.\ sep.: maximum angular separation between cluster members.
Flags: (a) switch between Gaussian and disk across IEM models.
(b) Fitted extension exceeds inner analysis radius ($2\times$ cluster size).
(c) Fitted position offset exceeds cluster size.
(d) Unreliable fit ($\mathrm{TS}_\mathrm{ext} \leq 0$, $\Gamma \geq 4$, or systematic uncertainty exceeds central value).}
\end{table}

\begin{table}[h!]
\caption{\label{tab:cluster_sources}Cluster members and 4FGL-DR4 classifications.}
\centering
\footnotesize
\setlength{\tabcolsep}{3pt}
\renewcommand{\arraystretch}{0.92}
\begin{tabular}{lp{0.78\textwidth}}
\hline\hline
ID & Source names (CLASS1) \\
\hline
1 & 4FGL J0616.5$+$2235, 4FGL J0617.2$+$2234e (SNR) \\
2 & 4FGL J0618.9$+$2240c, 4FGL J0620.1$+$2246 \\
3 & 4FGL J1111.8$-$6039 (PSR), 4FGL J1112.1$-$6108 (PSR), 4FGL J1112.2$-$6055 \\
4 & 4FGL J1320.5$-$6256c, 4FGL J1321.1$-$6239 \\
5 & 4FGL J1401.2$-$6116, 4FGL J1401.9$-$6130c \\
6 & 4FGL J1415.3$-$6110c, 4FGL J1416.4$-$6123c \\
7 & 4FGL J1420.3$-$6046e (PWN), 4FGL J1420.0$-$6048 (PSR), 4FGL J1418.7$-$6057 (PSR), 4FGL J1417.7$-$6057, 4FGL J1418.7$-$6110 \\
8 & 4FGL J1420.9$-$6127c, 4FGL J1422.5$-$6137 (PSR) \\
9 & 4FGL J1449.8$-$5923c, 4FGL J1450.2$-$5937c (spp) \\
10 & 4FGL J1510.1$-$5750, 4FGL J1511.2$-$5803 \\
11 & 4FGL J1514.1$-$5805c, 4FGL J1515.6$-$5817c \\
12 & 4FGL J1624.1$-$4941c, 4FGL J1622.7$-$4934c \\
13 & 4FGL J1633.0$-$4746e (spp), 4FGL J1634.0$-$4742c, 4FGL J1631.6$-$4756e (pwn) \\
14 & 4FGL J1636.9$-$4710c, 4FGL J1638.4$-$4715c \\
15 & 4FGL J1638.1$-$4641c, 4FGL J1638.5$-$4657c (spp), 4FGL J1639.8$-$4642c, 4FGL J1640.7$-$4631e (spp), 4FGL J1641.0$-$4619 (spp) \\
16 & 4FGL J1648.4$-$4611 (PSR), 4FGL J1648.4$-$4554 \\
17 & 4FGL J1711.9$-$4056c, 4FGL J1712.9$-$4105 \\
18 & 4FGL J1715.3$-$3832 (unk), 4FGL J1714.4$-$3830 (PSR), 4FGL J1714.8$-$3849 \\
19 & 4FGL J1729.5$-$3625c, 4FGL J1729.2$-$3641c (unk) \\
20 & 4FGL J1738.1$-$2453, 4FGL J1738.2$-$2510 \\
21 & 4FGL J1739.2$-$2717, 4FGL J1739.2$-$2732 \\
22 & 4FGL J1739.7$-$2836, 4FGL J1740.4$-$2850 \\
23 & 4FGL J1741.1$-$3328, 4FGL J1742.3$-$3318 (PSR) \\
24 & 4FGL J1743.5$-$2616, 4FGL J1744.5$-$2612 (unk) \\
25 & 4FGL J1747.2$-$2957 (PSR), 4FGL J1747.8$-$3006 \\
26 & 4FGL J1754.3$-$4443, 4FGL J1752.8$-$4449 \\
27 & 4FGL J1754.6$-$2933, 4FGL J1754.3$-$2915 \\
28 & 4FGL J1757.9$-$2419 (PSR), 4FGL J1758.6$-$2404 (unk) \\
29 & 4FGL J1758.8$-$2326, 4FGL J1759.5$-$2312 (unk) \\
30 & 4FGL J1759.1$-$2347c (unk), 4FGL J1759.7$-$2354, 4FGL J1800.1$-$2417c, 4FGL J1800.2$-$2403c, 4FGL J1800.7$-$2355 (unk), 4FGL J1800.9$-$2407, 4FGL J1801.8$-$2358 \\
31 & 4FGL J1801.3$-$2326e (SNR), 4FGL J1801.6$-$2326 \\
32 & 4FGL J1806.2$-$2126 (unk), 4FGL J1805.6$-$2136e (SNR), 4FGL J1804.7$-$2144e (SPP) \\
33 & 4FGL J1819.4$-$1102, 4FGL J1820.5$-$1059 \\
34 & 4FGL J1829.4$-$1500c, 4FGL J1830.0$-$1515c \\
35 & 4FGL J1837.8$-$0604 (PSR), 4FGL J1837.9$-$0620, 4FGL J1838.4$-$0630c, 4FGL J1838.4$-$0545, 4FGL J1838.7$-$0601, 4FGL J1838.9$-$0537 (PSR), 4FGL J1839.4$-$0553 (unk) \\
36 & 4FGL J1839.2$-$0420 (unk), 4FGL J1840.0$-$0411 (spp) \\
37 & 4FGL J1846.9$-$0247c, 4FGL J1846.4$-$0258 (pwn) \\
38 & 4FGL J1847.2$-$0141, 4FGL J1847.2$-$0200c, 4FGL J1847.7$-$0125, 4FGL J1848.1$-$0148c, 4FGL J1848.6$-$0202c (unk), 4FGL J1848.7$-$0129 (glc), 4FGL J1849.4$-$0117 \\
39 & 4FGL J1854.1$+$0142c, 4FGL J1854.7$+$0153, 4FGL J1855.8$+$0150 \\
40 & 4FGL J1855.2$+$0756, 4FGL J1856.2$+$0749 (unk) \\
41 & 4FGL J1856.7$+$0125c, 4FGL J1855.9$+$0121e (SNR) \\
42 & 4FGL J1857.4$+$0106, 4FGL J1857.1$+$0056 \\
43 & 4FGL J1859.2$+$0046, 4FGL J1859.3$+$0058 \\
44 & 4FGL J1906.4$+$0723 (PSR), 4FGL J1906.9$+$0712, 4FGL J1907.6$+$0703 (snr) \\
45 & 4FGL J1910.2$+$0904c, 4FGL J1911.0$+$0905 (snr), 4FGL J1911.8$+$0903c \\
46 & 4FGL J1925.2$+$1600c, 4FGL J1925.2$+$1618e (SNR) \\
47 & 4FGL J2240.3$-$5241, 4FGL J2241.7$-$5236 (MSP) \\
48 & 4FGL J2250.6$+$5809, 4FGL J2252.6$+$5808 (spp) \\
\hline
\end{tabular}
\tablefoot{CLASS1 classifications from 4FGL-DR4 are shown in parentheses where available.}
\end{table}

\twocolumn

\section{Clustering radius analysis}
\label{appendix:clustering_radius}

\begin{figure}[htbp]
\centering
\includegraphics[width=\columnwidth]{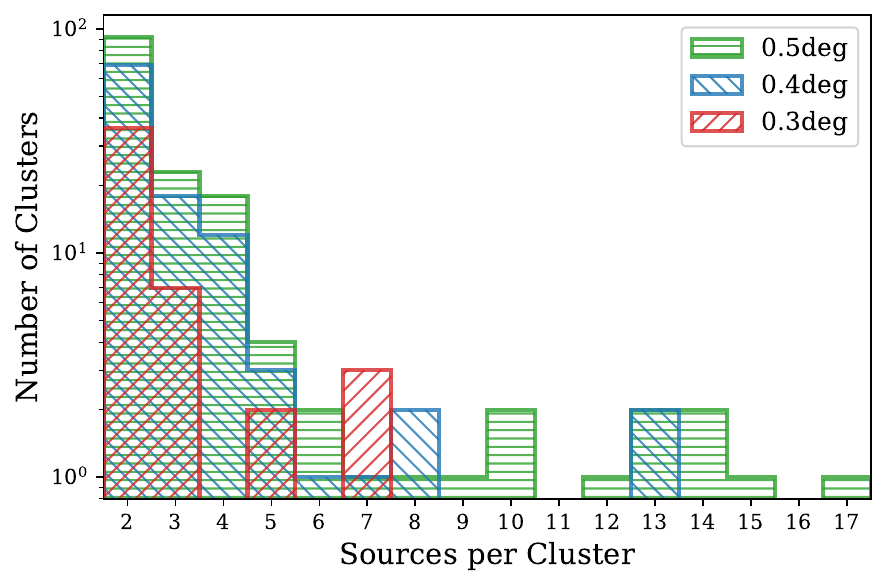}
\caption{Cluster size distributions for different DBSCAN clustering radii. The histograms show the number of clusters containing two or more sources for epsilon values of $0.3^\circ$, $0.4^\circ$, and $0.5^\circ$.}
\label{fig:clusters_size}
\end{figure}

We re-ran the DBSCAN algorithm with $\varepsilon = 0.3^\circ$, $0.4^\circ$, and $0.5^\circ$ to check how sensitive the clustering is to this choice. Table~\ref{tab:radius_analysis_summary} summarizes the statistics for each search radius. Larger radii naturally draw more sources into clusters (Table~\ref{tab:radius_analysis_summary}). The size distribution remains dominated by two- and three-source clusters at all radii (Fig.~\ref{fig:clusters_size}), though at $0.5^\circ$ the tail extends to clusters as large as 17 sources. At this scale the mean cluster radius is $0.818^\circ$ (Table~\ref{tab:radius_analysis_summary}), so large-scale diffuse-model patch templates on $\sim2^\circ$ scales become more relevant than for the nominal $0.3^\circ$ clustering.

\begin{figure}[htbp]
\centering
\includegraphics[width=\columnwidth]{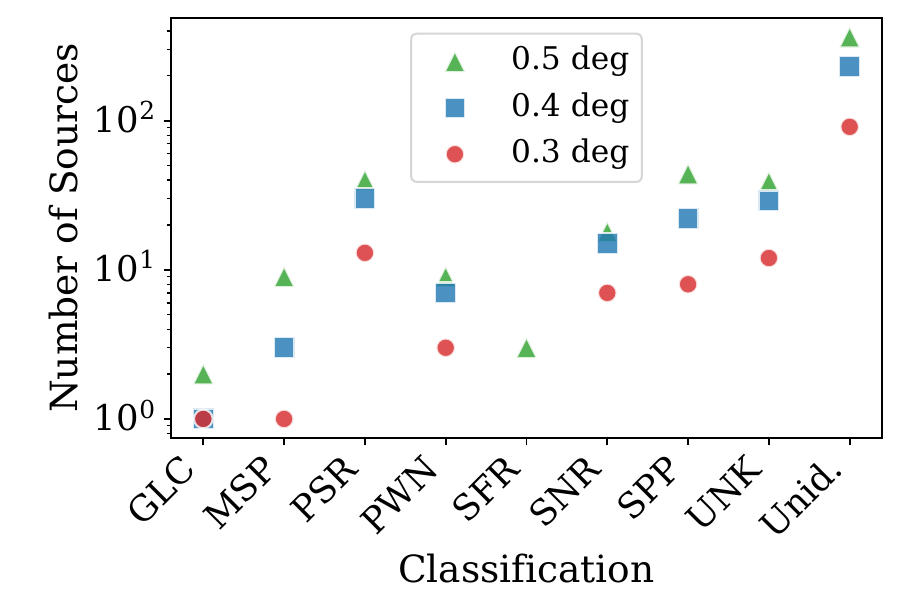}
\caption{Source classification distributions within clusters for different DBSCAN clustering radii. The histogram shows the composition of clustered sources by their 4FGL-DR4 catalog classifications (unassociated, PSR, SPP, SNR, etc.) for epsilon values of $0.3^\circ$, $0.4^\circ$, and $0.5^\circ$. Both firm identifications and associations from the 4FGL-DR4 \texttt{CLASS1} column are grouped under each class label. Missing symbols indicate that no sources of that class are present in any cluster at the corresponding search radius. In particular, no SFR-classified sources are present at $0.3^\circ$ or $0.4^\circ$. GLC-classified sources are present at all three radii, but the $0.3^\circ$ and $0.4^\circ$ values are both one and their markers overlap.}
\label{fig:sources_classification}
\end{figure}

Unassociated sources comprise the largest fraction of clustered sources across all search radii, while PSR and SPP represent the next most common classifications (see Fig.~\ref{fig:sources_classification}). Sources classified as UNK (unknown association) are as numerous as PSR and more frequent than SPP at $0.3^\circ$ and $0.4^\circ$. Like unassociated sources, their nature is uncertain and they may trace extended structures not yet firmly identified. SNR classifications are consistently represented across all radii, indicating that these extended objects are reliably recovered regardless of the clustering scale. At larger search radii ($0.5^\circ$), more sources classified as SPP, SNR, and PWN are drawn into clusters, as expected when the linking length exceeds the typical separation between catalog entries in crowded regions.

To understand how individual clusters evolve with changing search radius, we performed a cluster overlap analysis that reveals several patterns. When the search radius is increased from $0.3^\circ$ to $0.4^\circ$, 22 of the 48 clusters remain identical, which means their members are already separated by less than $0.3^\circ$ and the grouping is insensitive to the exact $\varepsilon$ choice. Another 26 clusters expand by picking up additional nearby sources, consistent with a hierarchical picture in which compact cores sit within more loosely associated surroundings. The remaining five cases are mergers, where two or more clusters at $0.3^\circ$ collapse into a single cluster at $0.4^\circ$. All five mergers occur in regions of high source density (e.g., the Kookaburra and W~28 fields), suggesting that the smaller-scale groupings are substructures of larger emission complexes.

Focusing on the 25 clusters that passed the initial selection (Table~\ref{tab:clusters_all}), seven retain identical membership at $0.4^\circ$ (Clusters~2, 4, 17, 23, 31, 39, and 43), while the remaining 18 expand to incorporate additional nearby sources. Notable examples include Clusters~6, 7, and 8 in the Kookaburra region, which all merge into a single large cluster ($0.4^\circ$ Cluster~24) containing 13 sources, and Clusters~28, 29, 30, and 31 in the W~28 region, which coalesce into Cluster~71 with 13 sources. At the $0.5^\circ$ scale, only one cluster (Cluster~17) maintains its identical structure, while all others expand substantially, some growing by factors of four to six in source count.

\begin{table}[ht]
\caption{Clustering analysis summary across different search radii.}
\label{tab:radius_analysis_summary}
\centering
\small
\begin{tabular}{lccc}
\hline\hline
Metric & $0.3^\circ$ & $0.4^\circ$ & $0.5^\circ$ \\
\hline
Total sources & 124 & 310 & 499 \\
Total clusters & 48 & 108 & 151 \\
Mean sources per cluster & 2.58 & 2.87 & 3.30 \\
Unassociated sources & 91 & 230 & 364 \\
Unassociated fraction (\%) & 73.4 & 74.2 & 72.9 \\
Mean cluster radius (deg) & 0.346 & 0.496 & 0.818 \\
Identical vs.\ previous & \dots & 22 & 56 \\
Expanded vs.\ previous & \dots & 26 & 52 \\
Merged vs.\ previous & \dots & 5 & 11 \\
\hline
\end{tabular}
\tablefoot{Mean sources per cluster indicates the average number of 4FGL sources per cluster. Identical, expanded, and merged refer to comparison with the next smaller search radius.}
\end{table}

\onecolumn

\section{TS maps and SEDs of significant clusters}
\label{appendix:tsmaps}

These maps correspond to results obtained with the standard Galactic diffuse emission model only. Results for alternative IEM models are not shown but were used to derive systematic uncertainties reported in Table~\ref{tab:clusters_fit_results}. The maps show the spatial distribution of gamma-ray emission significance across each cluster region. The TS maps obtained with the unpatched and \texttt{GALPROP} models were inspected for all clusters and are qualitatively consistent with those shown here. The quantitative differences are captured by the systematic uncertainties reported in Table~\ref{tab:clusters_fit_results}. Alongside each TS map, we show the SED of the extended-source model for the standard IEM. Upper limits are shown for flux points with significance lower than $2\sigma$. The maps and SEDs for the selected clusters are shown in Fig.~\ref{fig:tsmaps1}.

\begin{figure*}[h!]
    \centering
    \captionsetup[subfigure]{skip=0pt}
    \begin{subfigure}[b]{0.45\textwidth}
        \centering
        \includegraphics[width=\textwidth]{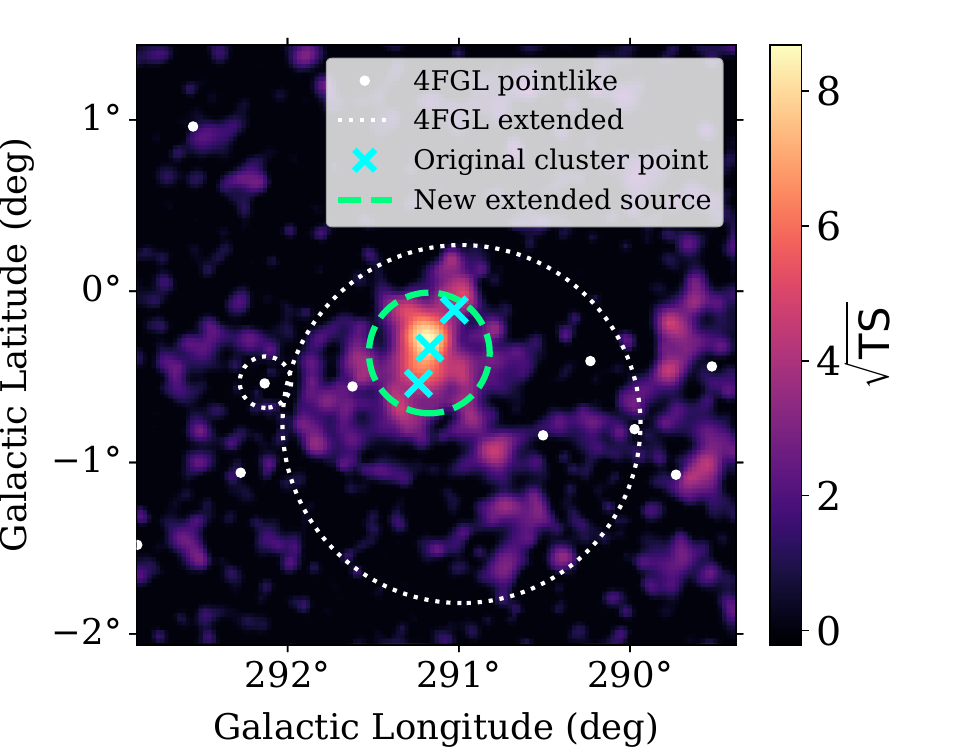}
        \caption*{Cluster 3, TS map}
    \end{subfigure}
    \hspace{0.5cm}
    \begin{subfigure}[b]{0.457\textwidth}
        \centering
        \includegraphics[width=\textwidth]{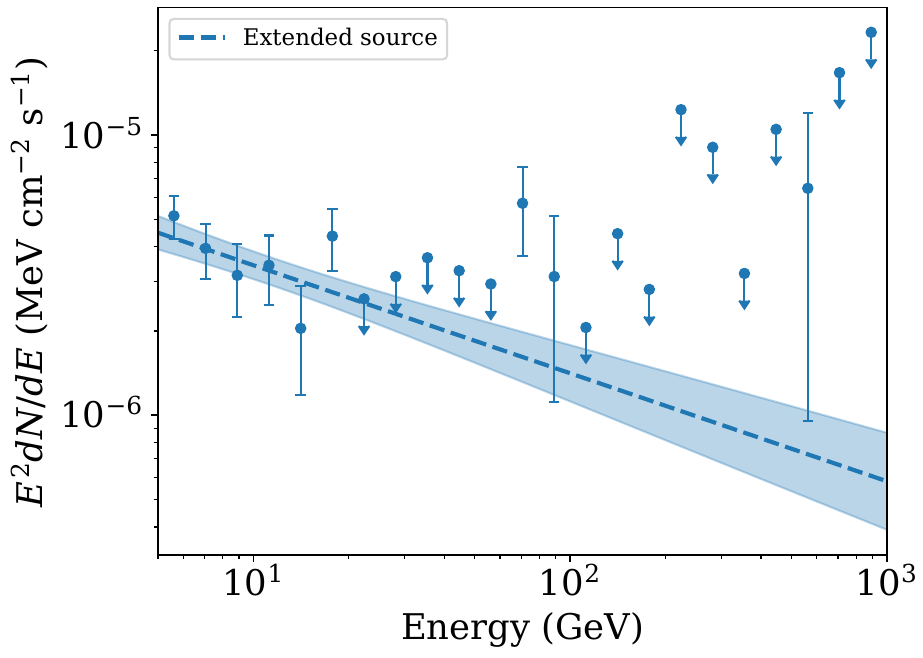}
        \caption*{Cluster 3, SED}
    \end{subfigure}
    \vspace{-0.10cm}
    \begin{subfigure}[b]{0.45\textwidth}
        \centering
        \includegraphics[width=\textwidth]{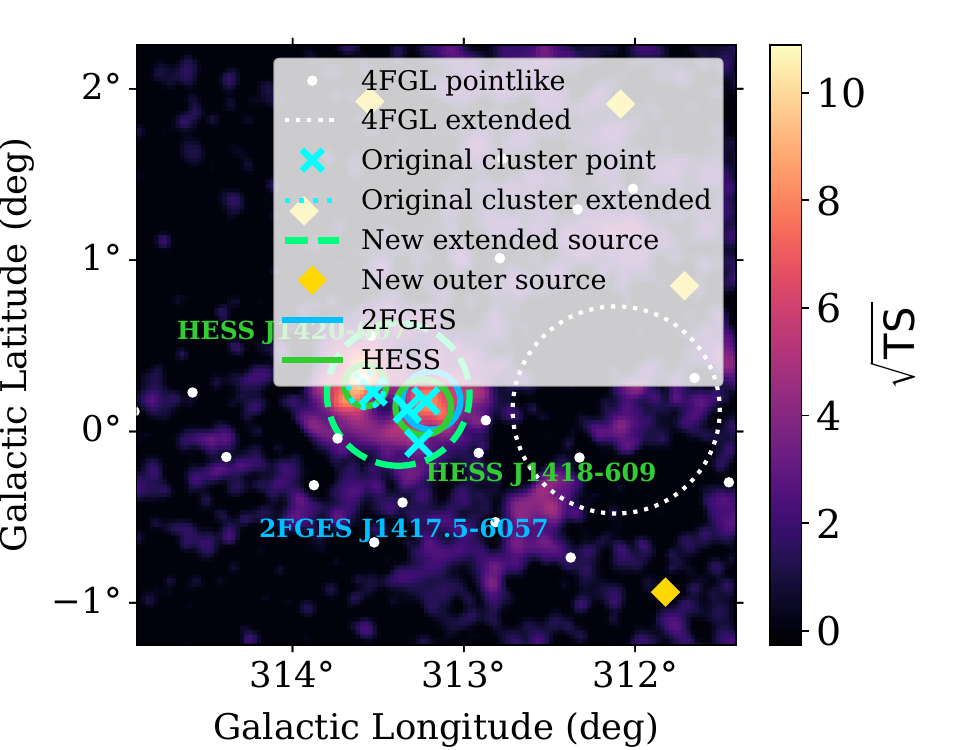}
        \caption*{Cluster 7, TS map}
    \end{subfigure}
    \hspace{0.5cm}
    \begin{subfigure}[b]{0.457\textwidth}
        \centering
        \includegraphics[width=\textwidth]{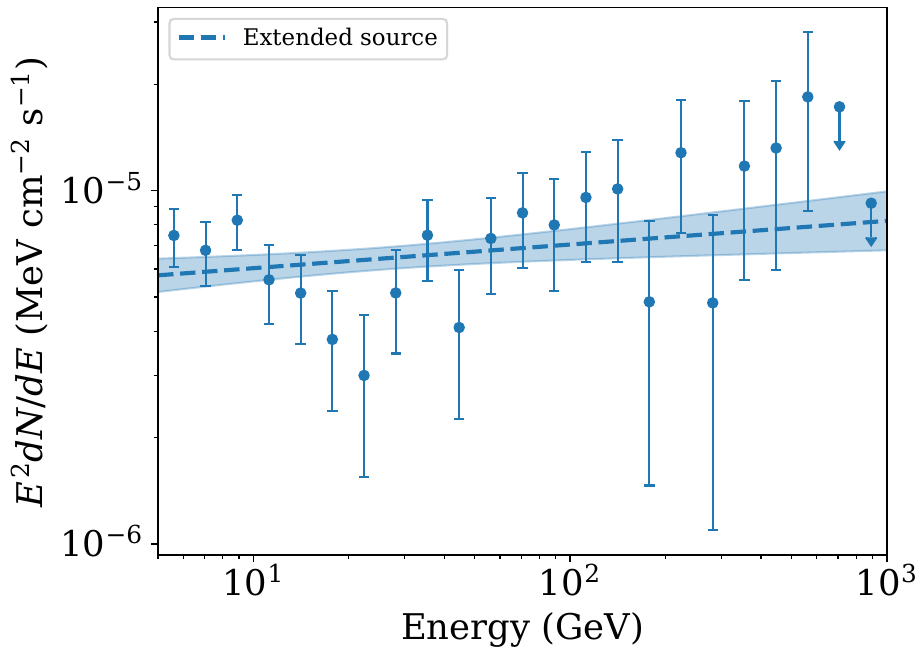}
        \caption*{Cluster 7, SED}
    \end{subfigure}
    \vspace{-0.10cm}
    \begin{subfigure}[b]{0.45\textwidth}
        \centering
        \includegraphics[width=\textwidth]{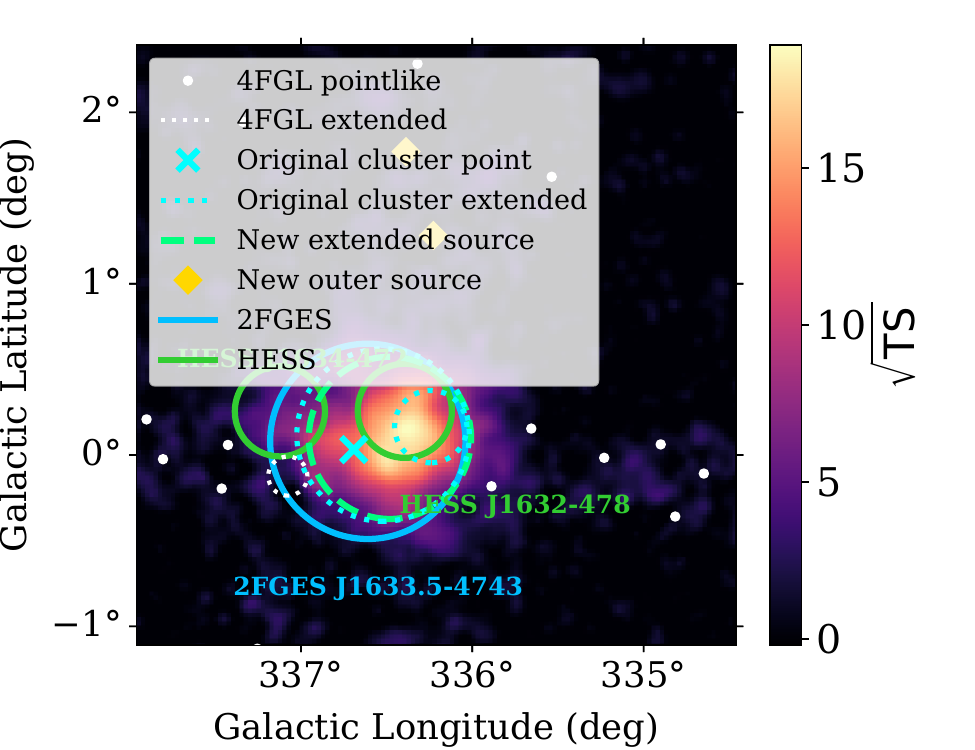}
        \caption*{Cluster 13, TS map}
    \end{subfigure}
    \hspace{0.5cm}
    \begin{subfigure}[b]{0.457\textwidth}
        \centering
        \includegraphics[width=\textwidth]{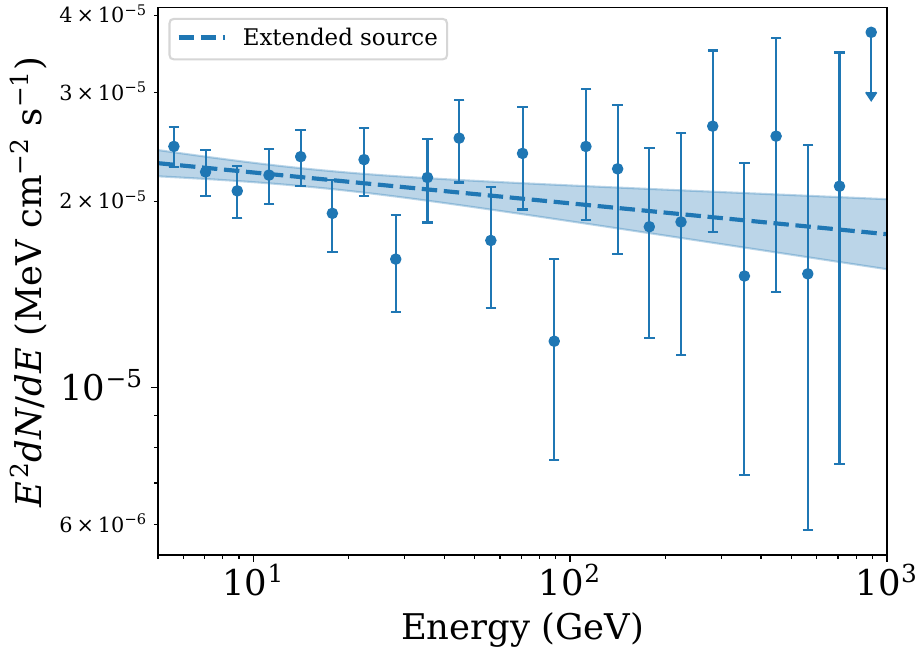}
        \caption*{Cluster 13, SED}
    \end{subfigure}
    \caption{TS maps (left) and SEDs (right) for clusters 3, 7, 13.}
    \label{fig:tsmaps1}
\end{figure*}

\begin{figure*}[h!]
    \ContinuedFloat
    \centering
    \captionsetup[subfigure]{skip=0pt}
    \begin{subfigure}[b]{0.45\textwidth}
        \centering
        \includegraphics[width=\textwidth]{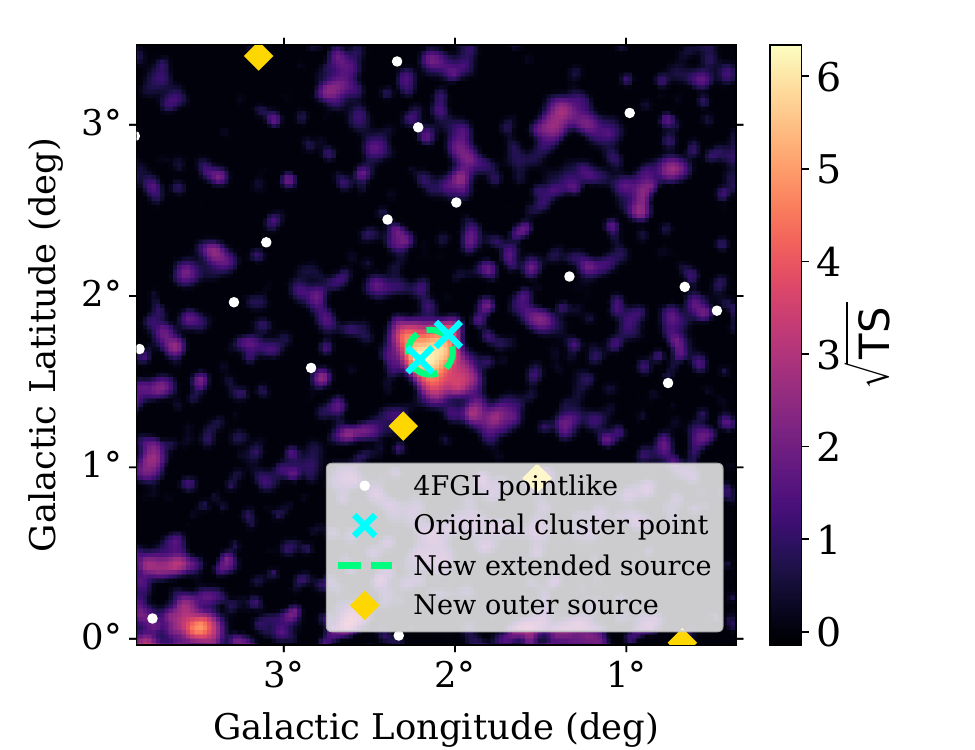}
        \caption*{Cluster 24, TS map}
    \end{subfigure}
    \hspace{0.5cm}
    \begin{subfigure}[b]{0.457\textwidth}
        \centering
        \includegraphics[width=\textwidth]{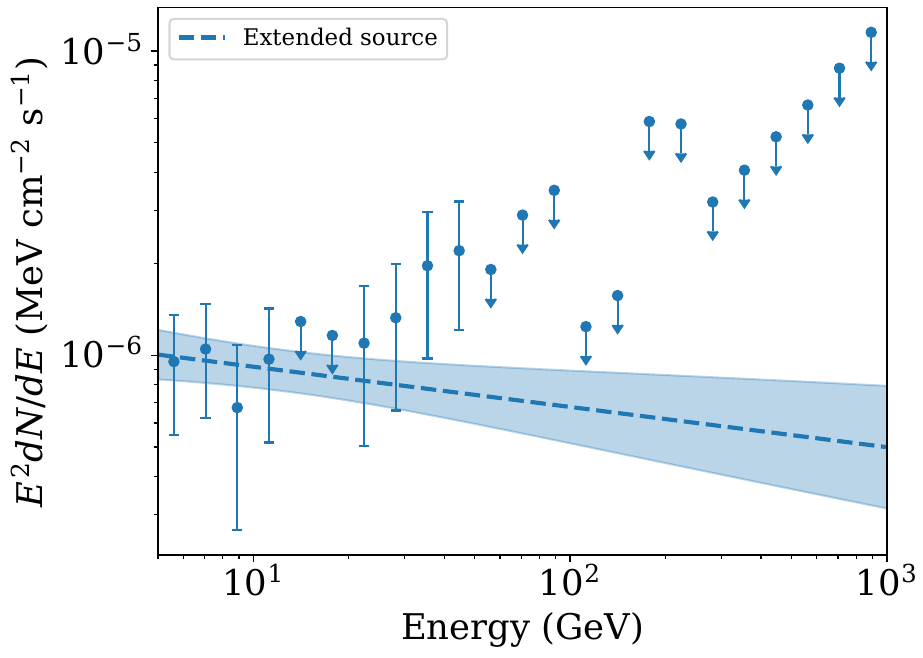}
        \caption*{Cluster 24, SED}
    \end{subfigure}
    \vspace{-0.10cm}
    \begin{subfigure}[b]{0.45\textwidth}
        \centering
        \includegraphics[width=\textwidth]{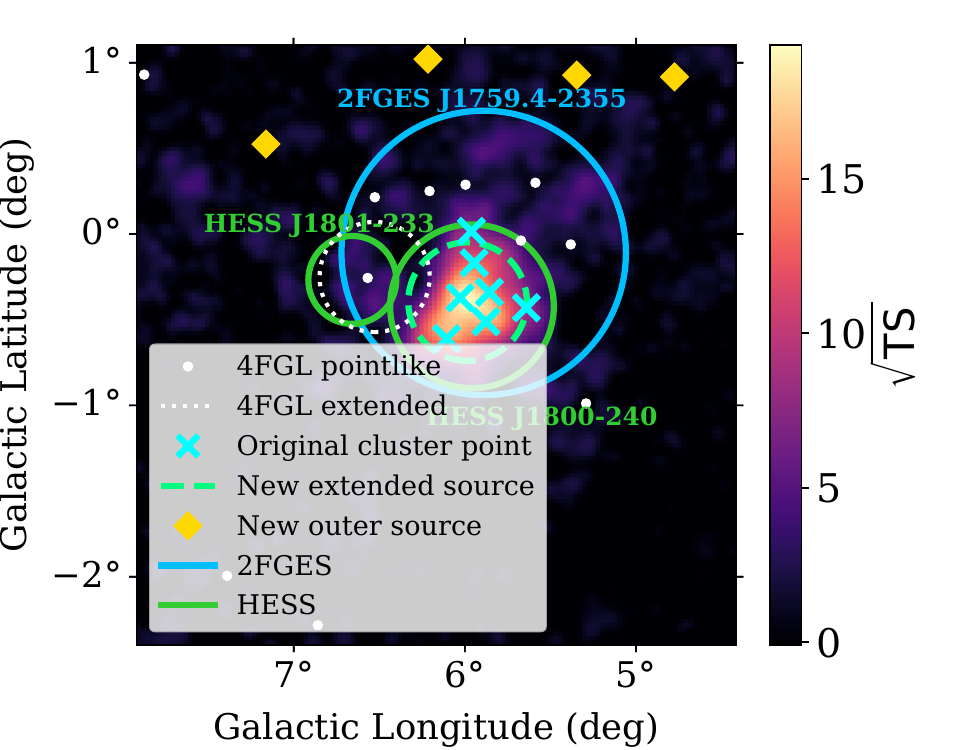}
        \caption*{Cluster 30, TS map}
    \end{subfigure}
    \hspace{0.5cm}
    \begin{subfigure}[b]{0.457\textwidth}
        \centering
        \includegraphics[width=\textwidth]{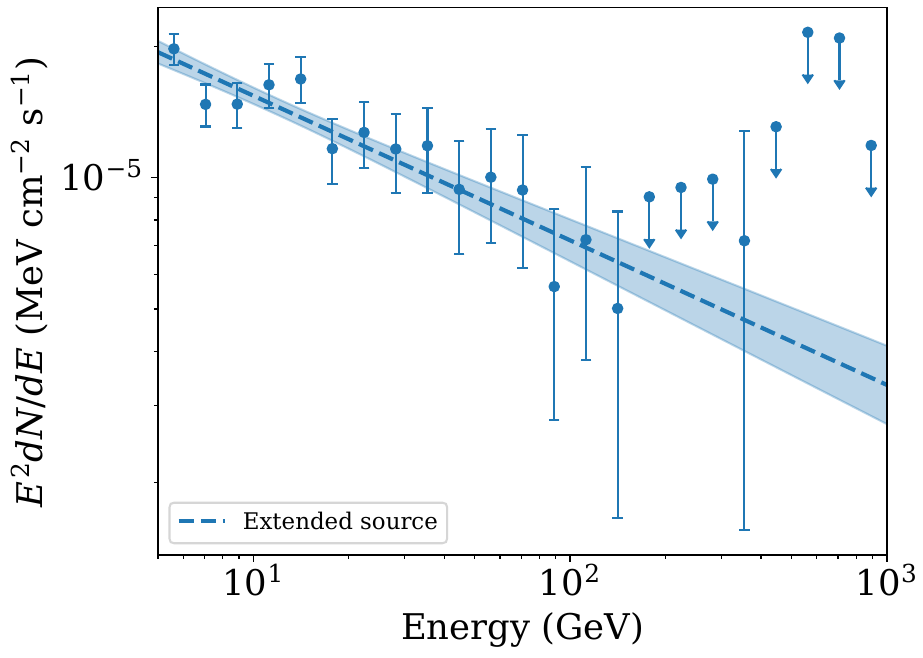}
        \caption*{Cluster 30, SED}
    \end{subfigure}
    \vspace{-0.10cm}
    \begin{subfigure}[b]{0.45\textwidth}
        \centering
        \includegraphics[width=\textwidth]{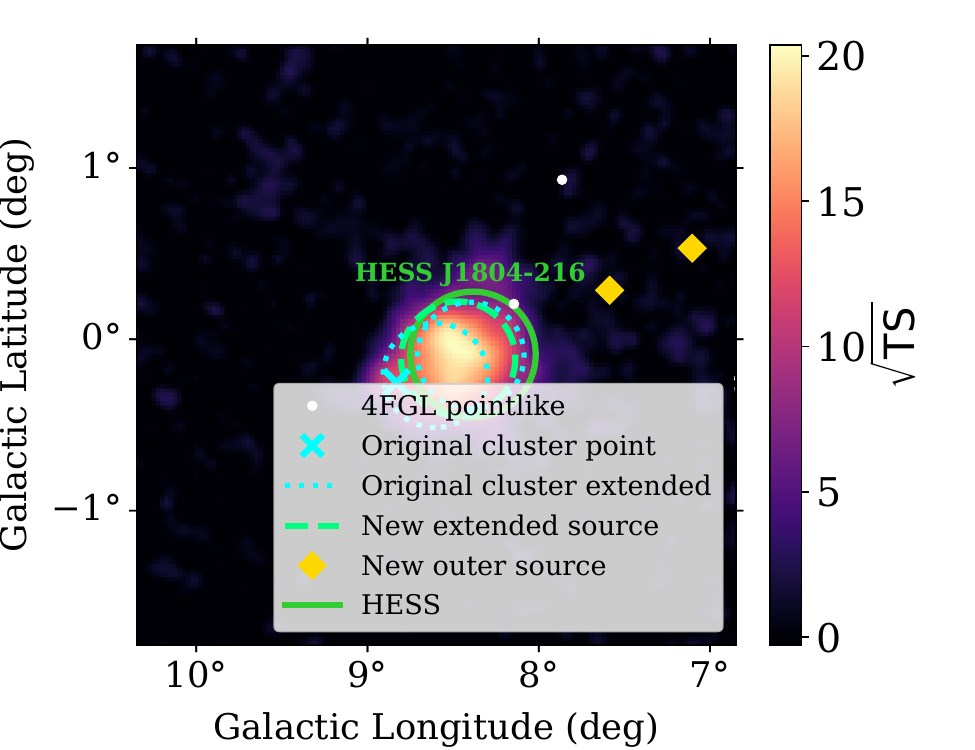}
        \caption*{Cluster 32, TS map}
    \end{subfigure}
    \hspace{0.5cm}
    \begin{subfigure}[b]{0.457\textwidth}
        \centering
        \includegraphics[width=\textwidth]{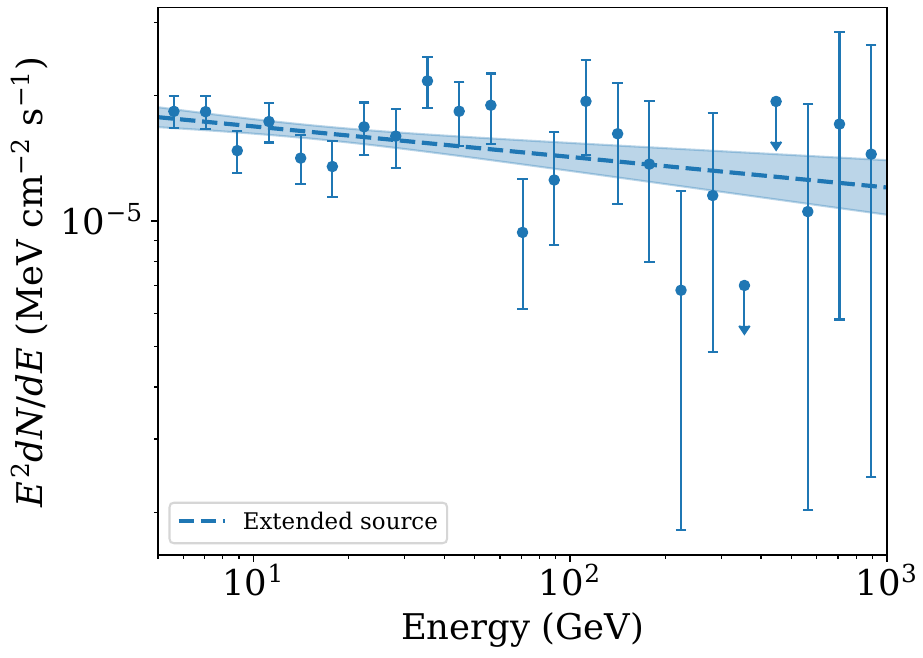}
        \caption*{Cluster 32, SED}
    \end{subfigure}
    \caption{TS maps (left) and SEDs (right) for clusters 24, 30, 32.}
    \label{fig:tsmaps2}
\end{figure*}

\begin{figure*}[h!]
    \ContinuedFloat
    \centering
    \captionsetup[subfigure]{skip=0pt}
    \begin{subfigure}[b]{0.45\textwidth}
        \centering
        \includegraphics[width=\textwidth]{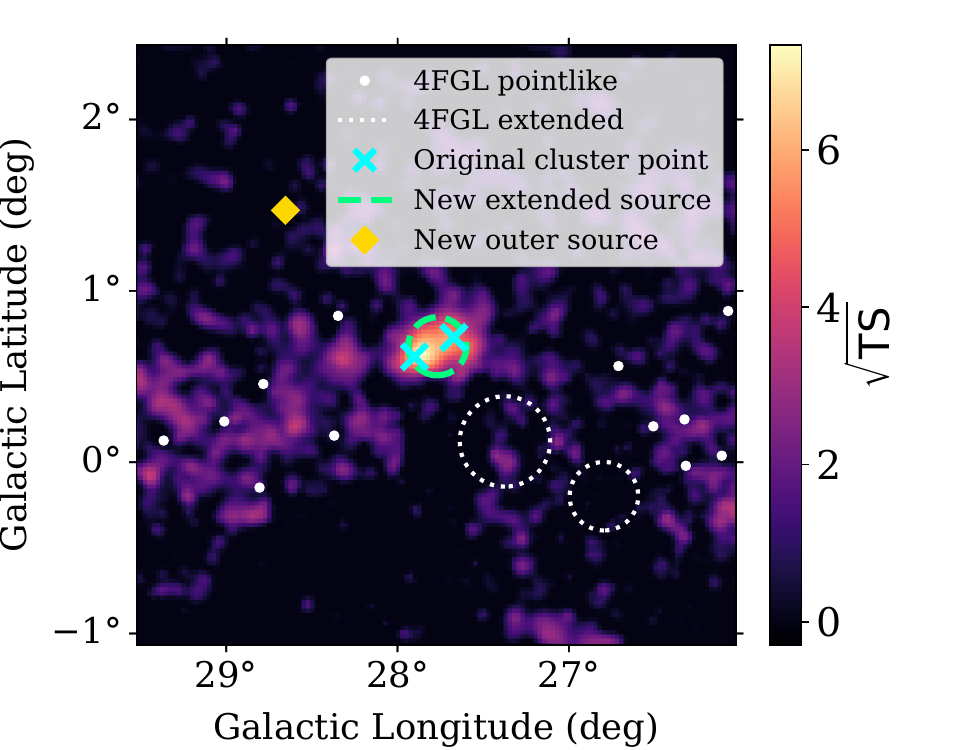}
        \caption*{Cluster 36, TS map}
    \end{subfigure}
    \hspace{0.5cm}
    \begin{subfigure}[b]{0.457\textwidth}
        \centering
        \includegraphics[width=\textwidth]{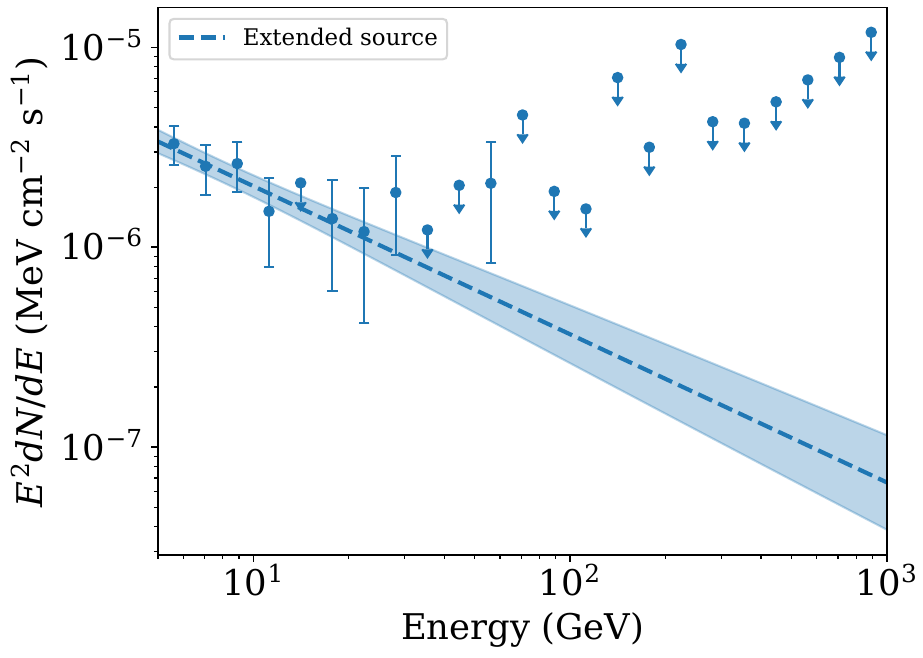}
        \caption*{Cluster 36, SED}
    \end{subfigure}
    \vspace{-0.10cm}
    \begin{subfigure}[b]{0.45\textwidth}
        \centering
        \includegraphics[width=\textwidth]{images/ts_maps/cluster_42_standard_null_ts_map.pdf}
        \caption*{Cluster 42, TS map}
    \end{subfigure}
    \hspace{0.5cm}
    \begin{subfigure}[b]{0.457\textwidth}
        \centering
        \includegraphics[width=\textwidth]{images/sed_plots/sed_42_standard.pdf}
        \caption*{Cluster 42, SED}
    \end{subfigure}
    \caption{TS maps (left) and SEDs (right) for clusters 36, 42.}
    \label{fig:tsmaps3}
\end{figure*}

\end{appendix}

\end{document}